\documentclass[conference]{IEEEtran}
\IEEEoverridecommandlockouts

\usepackage{algorithmic}
\usepackage[linesnumbered,ruled,vlined]{algorithm2e}
\usepackage{multirow}
\usepackage{multicol}
\usepackage{wrapfig}
\usepackage{soul}
\usepackage{balance}
\usepackage{epstopdf}

\usepackage{amsfonts}
\usepackage[flushleft]{threeparttable}
\usepackage{adjustbox}
\usepackage{fancyhdr}
\usepackage[dvipsnames]{xcolor}
\usepackage[flushleft]{threeparttable}
\usepackage{tikz}
\usepackage{outlines}
\usepackage{tcolorbox}
\usepackage{blindtext}
\usepackage{xcolor}
\usepackage{color}
\usepackage{listings}
\usepackage{color}
\usepackage{MnSymbol,wasysym}
\usepackage[switch]{lineno}
\usepackage{array}
\usepackage{caption}

\usepackage{amsmath,amsthm}

\usepackage{mathtools} 

\usepackage{subfigure}
\usepackage{subfig}

\usepackage{xcolor,colortbl}
\def\BibTeX{{\rm B\kern-.05em{\sc i\kern-.025em b}\kern-.08em
    T\kern-.1667em\lower.7ex\hbox{E}\kern-.125emX}}

\newcommand{\rulesep}{\unskip\ \vrule\ }

\usepackage{tikz}
\newcommand*\circled[1]{\tikz[baseline=(char.base)]{
            \node[shape=circle,draw,inner sep=2pt] (char) {#1};}}

\begin{document}

\title{Dynasor: A \underline{Dyna}mic Memory Layout for Accelerating Sparse MTTKRP for Ten\underline{sor} Decomposition on Multi-core CPU}

\author{
    \IEEEauthorblockN{Sasindu Wijeratne\IEEEauthorrefmark{1}, Rajgopal Kannan\IEEEauthorrefmark{2}, Viktor Prasanna\IEEEauthorrefmark{1}}
    \IEEEauthorblockA{\IEEEauthorrefmark{1} University of Southern California, Los Angeles, USA}
    \IEEEauthorblockA{\IEEEauthorrefmark{2} DEVCOM Army Research Lab, Los Angeles, USA}
    Email: \{kangaram, prasanna\}@usc.edu, rajgopal.kannan.civ@army.mil
}


\maketitle

\begin{abstract}
Sparse Matricized Tensor Times Khatri-Rao Product (spMTTKRP) is the most time-consuming compute kernel in sparse tensor decomposition. In this paper, we introduce a novel algorithm to minimize the execution time of spMTTKRP across all modes of an input tensor on multi-core CPU platform. The proposed algorithm leverages the FLYCOO tensor format to exploit data locality in external memory accesses. It effectively utilizes computational resources by enabling lock-free concurrent processing of independent partitions of the input tensor. The proposed partitioning ensures load balancing among CPU threads. Our dynamic tensor remapping technique leads to reduced communication overhead along all the modes. On widely used real-world tensors, our work achieves 2.12$\times$ - 9.01$\times$ speedup in total execution time across all modes compared with the state-of-the-art CPU implementations.
\end{abstract}


\vspace{-3mm}
\section{Introduction}
Tensor Decomposition (TD) enables the transformation of high-dimensional tensors into a lower-dimensional latent space, facilitating the identification of important features in the data distribution. TD finds applications in various fields such as machine learning~\cite{7891546, mondelli2019connection, cheng2020novel}, signal processing~\cite{wen2020tensor}, and network analysis~\cite{fernandes2020tensor}. Canonical Polyadic Decomposition (CPD) via alternating least squares (CP-ALS) is a widely used TD algorithm where Matricized Tensor Times Khatri-Rao Product (MTTKRP) is the most time-consuming computation. 


Numerous tensor formats have been proposed in the literature to tackle the challenges posed by sparse tensors encountered in real-world scenarios~\cite{10.1145/3543622.3573179, alto_paper, 8665782, 7161496, 10.1145/3295500.3356216, 10.1145/3330345.3330366}. These formats employ multiple tensor copies (i.e., mode-specific tensor formats) or additional memory to store intermediate results of computations to support irregular data access patterns in each tensor mode. However, these approaches have the drawback of increasing the overall memory requirements of the algorithm. Notably, mode-specific tensor formats necessitate several replicas of the original tensor, each arranged based on different permutations of nonzero tensor elements. This replication grows linearly with the number of modes, rendering it impractical as the number of modes increases. Additionally, relying on memory to store intermediate values poses scalability challenges, limiting its usefulness to small datasets. Furthermore, as the size of the tensor grows, there is a potential for memory explosion, further exacerbating the scalability issue.

One desirable solution to reduce memory traffic is to reduce the number of accesses to the data by increasing their reusability. 
Tensor formats such as HiCOO~\cite{8665782} and ALTO~\cite{alto_paper}, which use variations of Morton ordering~\cite{8665782} to bring the tensor elements with neighboring coordinates closer, exhibit an increase of data reusability. However, these tensor formats still generate a significant number of intermediate values that usually lead to increased memory access time and storage requirements.

Wijeratne et al.~\cite{10.1145/3543622.3573179} introduced FLYCOO, a novel tensor format aimed at accelerating spMTTKRP on Field Programmable Gate Arrays (FPGAs). FLYCOO enhances data locality across all tensor modes while accessing the input tensor and the factor matrices in the FPGA external memory. 

We employ a dynamic memory layout to achieve load balancing and communication efficiency, enabling a straightforward static schedule for computing spMTTKRP in each mode. This dynamic approach significantly reduces the generation of intermediate values during computation, which would otherwise need to be communicated to the CPU external memory. To facilitate the memory layout, we propose a parallel algorithm, Dynasor, which performs elementwise computations and dynamic tensor remapping.







The key contributions of this work are:

\begin{itemize}



\item We adapt the FLYCOO data format to support spMTTKRP computation on multi-core CPUs by introducing, Dynasor: a parallel algorithm for spMTTKRP with dynamic tensor remapping. We implement Dynasor using C++ and openMP directives.

\item \textbf{Dynamic memory layout:} By dynamically remapping the input tensor between successive executions of spMTTKRP, we increase the data locality of external memory accesses while avoiding the intermediate value communication to the external memory. Our empirical results demonstrate that our approach can perform under a memory-constrained environment with minimum external memory during runtime compared to state-of-the-art implementations when applied to large tensors.

\item \textbf{Scheduling and Load balancing:} 
Our proposed Super-shard-based load balancing technique enables lock-free spMTTKRP computation. Compared to the single-thread implementation, the paper introduces a static scheduling scheme that achieves a speedup of 8.5$\times$ to 21$\times$ on 56 CPU threads.





\item On widely used real-world tensors, our work achieves 2.12$\times$ - 9.01$\times$ speedup in total execution time across all modes compared to the state-of-the-art CPU implementations.


\end{itemize}

\section{Background and Related Work}\label{background}
\subsection{Notations}
Table~\ref{table:notation} summarizes the notations used in this paper.

\vspace{-3mm}
\begin{table}[ht]
\caption{Notations}
\vspace{-5mm}
\begin{center}
\begin{tabular}{c|l}
     Symbol & Details \\
     \hline
     $\mathcal{X}$ & sparse tensor \\
     $\mathcal{X}_{(n)}$ & mode-$n$ matricization of $\mathcal{X}$ \\
$\mathbf{M}$ & matrix \\
$\mathbf{v}$ & vector \\
     $a$ & scalar \\
          $\circ$ & vector outer product \\
     $\otimes$ & Kronecker product \\
     $\odot$ & Khatri-Rao product \\
     \hline
\end{tabular}
\label{table:notation}
\end{center}
\end{table}
\vspace{-7mm}
\subsection{Tensor Decomposition}\label{background:decomp}

A tensor is a generalization of an array in multiple dimensions. In TD, the number of dimensions of an input tensor is commonly called the number of tensor modes. A $N$-mode, real-valued tensor is denoted by $\mathcal{X} \in \mathbb{R}^{I_0 \times \cdots \times I_{N-1}}$. Further, $\mathcal{X}_{(n)}$ denotes the mode-$n$ matricization or matrix unfolding~\cite{favier2014overview} of $\mathcal{X}$. $\mathcal{X}_{(n)}$ is defined as the matrix $\mathcal{X}_{(n)} \in \mathbb{R}^{I_n \times (I_0 \cdots I_{n-1} I_{n+1} \cdots I_{N-1})}$ where the parenthetical ordering indicates, the mode-$n$ column vectors are arranged by sweeping all the other mode indices through their ranges.

Canonical Polyadic Decomposition (CPD) decomposes $\mathcal{X}$ into a sum of single-mode tensors (i.e., arrays), which best approximates $\mathcal{X}$. For example, given 3-mode tensor $\mathcal{X} \in \mathbb{R}^{I_0 \times I_1 \times I_2}$, our goal is to approximate the original tensor as
\vspace{-3mm}
\begin{equation} \label{eqn_approx_tensor}
\begin{split}
\mathcal{X} \approx \sum_{r=0}^{R-1} \mathbf{a}_r \circ \mathbf{b}_r \circ \mathbf{c}_r
\end{split}
\end{equation}
\vspace{-3mm}

where $R$ is a positive integer and $\mathbf{a}_r \in \mathbb{R}^{I_0}$, $\mathbf{b}_r \in \mathbb{R}^{I_1}$, and $\mathbf{c}_r \in \mathbb{R}^{I_2}$. For a thorough review of CPD, refer to~\cite{kolda2009tensor}.

In the rest of Section~\ref{background}, we assume that the number of modes is three for illustration purposes.


\vspace{-5mm}
\begin{algorithm}
\DontPrintSemicolon
Input: A tensor $\mathcal{X} \in \mathbb{R}^{I_0 \times I_1 \times I_2}$, the rank $R \in \mathbb{Z}^{+}$ \;
Output: CP decomposition $[\![ \mathbf{A}, \mathbf{B}, \mathbf{C} ]\!]$, $\mathbf{A} \in \mathbb{R}^{I_0 \times R}$, $\mathbf{B} \in \mathbb{R}^{I_1 \times R}$, $\mathbf{C} \in \mathbb{R}^{I_2 \times R}$ \;
\While{\emph{stopping criterion not met}}{
\textcolor{blue}{// Matricization of $\mathcal{X}$ is different for each factor matrix computation} \;

    $\mathbf{A} \gets \mathbf{spMTTKRP}(\mathcal{X}_{(0)}, \mathbf{B}, \mathbf{C})$ \;
    $\mathbf{B} \gets \mathbf{spMTTKRP}(\mathcal{X}_{(1)}, \mathbf{A}, \mathbf{C})$ \;
    $\mathbf{C} \gets \mathbf{spMTTKRP}(\mathcal{X}_{(2)}, \mathbf{A}, \mathbf{B})$ \;
   Normalize $\mathbf{A}$, $\mathbf{B}$, $\mathbf{C}$ \;
}
\caption{CP-ALS for 3-mode tensors}
\label{cp-als}
\end{algorithm}

The alternating least squares (ALS) method is used to compute CPD. Algorithm~\ref{cp-als} shows the ALS method for CPD (i.e., CP-ALS) where Matricized Tensor-Times Khatri-Rao product (MTTKRP) is iteratively performed on all the Matricizations of $\mathcal{X}$, iteratively. In this paper, performing MTTKRP on all the Matricizations of an input tensor is called computing MTTKRP along all the modes. The outputs $\mathbf{A}$, $\mathbf{B}$, and $\mathbf{C}$ are the factor matrices that approximate $\mathcal{X}$. $\mathbf{a}_r$, $\mathbf{b}_r$, and $\mathbf{c}_r$ in Equation~\ref{eqn_approx_tensor} refers to the $r^{\text{th}}$ column of $\mathbf{A}$, $\mathbf{B}$, and $\mathbf{C}$, respectively. 

In this paper, we focus on MTTKRP on sparse tensors (spMTTKRP), which means the tensor is sparse, and the factor matrices are dense.


\vspace{-3mm}
\subsection{Elementwise computation of spMTTKRP}\label{sec_elementwise_compute}

The objective of this paper is to reduce the total execution time of spMTTKRP along all the modes of the tensor. The efficient execution of elementwise computation of spMTTKRP in all the tensor modes is the key to reducing the total execution time.

\begin{wrapfigure}{r}{0.28\textwidth}
\vspace{-5mm}
  \begin{center}
    \includegraphics[width=0.28\textwidth]{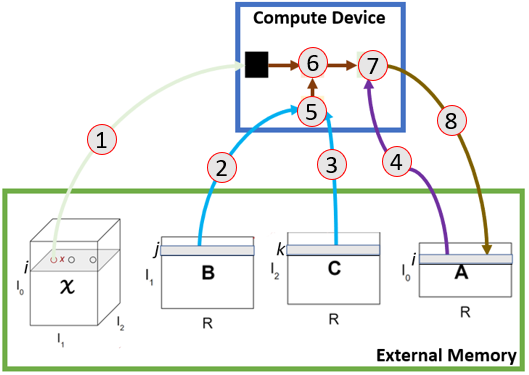}
  \end{center}
  \vspace{-4mm}
  \caption{Elementwise computation of spMTTKRP}
  \vspace{-3mm}
  \label{element_fig}
\end{wrapfigure}

Figure~\ref{element_fig} summarizes the elementwise computation of a nonzero tensor element in mode 0 of a 3-mode tensor. Here, we use the same notations as Algorithm~\ref{cp-als}.

In Figure~\ref{element_fig}, the elementwise computation is carried out on a nonzero tensor element, denoted as $\mathcal{X}_{(0)}(i,j,k)$. In sparse tensors, $\mathcal{X}_{(0)}(i,j,k)$ is typically represented in formats such as COOrdinate (COO) or Compressed Sparse Fiber (CSF). These formats store the indices ($i$, $j$, and $k$) or pointers to these indices along with the element value (i.e., $val(\mathcal{X}_{(0)}(i,j,k))$).

To perform the computation, $\mathcal{X}_{(0)}(i,j,k)$ is first loaded onto the Compute device from the external memory (step \circled{1}). The Compute device retrieves the rows $\mathbf{{A}}(i,:)$, $\mathbf{{B}}(j,:)$, and $\mathbf{{C}}(k,:)$ from the factor matrices using the index values extracted from $\mathcal{X}_{(0)}(i,j,k)$ (step \circled{2}, step \circled{3}, and step \circled{4}). Then the Compute device performs the following computation:
\[
\mathbf{{A}}(i,r) = \mathbf{{A}}(i,r) + val(\mathcal{X}_{(0)}(i,j,k)) \cdot \mathbf{{B}}(j,r) \cdot \mathbf{{C}}(k,r)
\]
Here, $r$ refers to the column index of a factor matrix row ($r < R$). The operation involves performing a Hadamard product between row $\mathbf{{B}}(j,:)$ and row $\mathbf{{C}}(j,:)$ (step \circled{5}), and then multiplying each element of the resulting product by $val(\mathcal{X}_{(0)}(i,j,k))$ (step \circled{6}). After updating $\mathbf{{A}}(i,:)$ (step \circled{7}), the updated value is stored back in the external memory (step \circled{8}).

\vspace{-2mm}
\subsection{Related Work}


Wijeratne et al.~\cite{10.1145/3543622.3573179} develop a customized accelerator on Field Programmable Gate Array (FPGA) for performing spMTTKRP on sparse tensors along with a specific tensor format labeled FLYCOO that supports the FPGA optimizations. In this paper, we focus on optimizing spMTTKRP for multi-core CPUs, which pose substantially different challenges from FPGAs, and adapt the FLYCOO format to support our multi-core CPU optimizations.


Helal et al.~\cite{alto_paper} propose ALTO, a tensor ordering method based on space-filling curves designed to effectively encode irregularly shaped spaces. ALTO requires minimum external memory to store tensors. However, this tensor format entails storing a large number of intermediate values generated during the computation in external memory during runtime, increasing memory access time. In contrast to ALTO, we use a dynamic memory layout to mitigate the communication time between the CPU and external memory.

J. Li et al.~\cite{8665782} propose HiCOO, a block-based format that utilizes compression techniques to handle sparse tensors by leveraging the Z-Morton curve~\cite{10.1007/3-540-44520-X_108} for efficient storage and retrieval. However, HiCOO encounters workload imbalance issues among blocks due to the irregular spatial distribution of sparse data. Despite both FLYCOO and HiCOO~\cite{8665782} employing similar tensor ordering strategies (i.e., Z-Morton ordering) during format generation, they demonstrate notable distinctions in terms of reduced intermediate value communication, tensor partitioning scheme, and distribution of nonzero elements. An in-depth comparison can be found in~\cite{10.1145/3543622.3573179}.

Kurt et al.~\cite{9820702} propose the STeF format to explore the impact of saving partial spMTTKRP results and reusing them during the spMTTKRP computation along all the tensor modes. They present a load-balancing approach at a fine-grained level to enable higher levels of parallelization. In contrast, our work employs a dynamic tensor remapping technique to optimize data locality during elementwise computation. Our proposed parallel algorithm also eliminates the need for additional storage to store partial spMTTKRP results, as required in the algorithm suggested by Kurt et al.~\cite{9820702}.
\section{FLYCOO Tensor Format} \label{hyper-graph} \label{Data_partitioning}
The FLYCOO tensor format is introduced in~\cite{10.1145/3543622.3573179} to perform spMTTKRP for tensor decomposition on FPGAs. Our work adapts the FLYCOO format to accelerate spMTTKRP on multi-core CPU. Section~\ref{Tensor_Format_Definition} provides a brief overview of the FLYCOO format. More details on FLYCOO format can be found in~\cite{10.1145/3543622.3573179}.

In tensor decomposition, spMTTKRP is computed along each mode sequentially as described in Section~\ref{background:decomp}. When computing spMTTKRP for mode $n$ of the input tensor, mode $n$ is referred to as the output mode and its corresponding factor matrix as the output factor matrix. Meanwhile, the rest of the modes are called input modes, and their factor matrices become input factor matrices.

The FLYCOO format assigns each nonzero tensor element to a tensor partition for each mode and embeds partition IDs to each tensor element. The tensor is divided into multiple partitions, called \textit{super-shards} with an equal number of output mode indices.

\subsection{FLYCOO Tensor Format} \label{Tensor_Format_Definition}
For each output mode $n$ ($0 \le n < N$), consider $m_n$ rows of the output factor matrix. Here, we are considering an input tensor with $N$-modes. Let $I_n$ denote the indices in mode $n$. FLYCOO reorganizes the data in the following manner: First, the indices of mode $n$ are partitioned into $k_n = \frac{|I_n|}{m_n}$ equal-sized sets $I_{n,0}, I_{n,1}, \ldots, I_{n, k_{n - 1}}$, where$|I_n|$ is the number of indices in mode $n$.
Next, the nonzero tensor elements incident on $I_{n, j}$ are collected into a {\it super-shard} $\textit{SS}_{n,j}$. Finally, to support dynamic tensor remapping, each super-shard is further divided into equal-sized sets of size $g$ called shards, where $g$ is a tensor partitioning parameter that is tuned depending on the cache size of the CPU platform (see Section~\ref{parameter_cpu}). The $q^{th}$ such shard is denoted as $shard_{n,j,q}$ and the total number of shards for mode $n$ is equal to $\sum_{j=0}^{k_n-1} \left\lceil|\textit{SS}_{n,j}|/g\right\rceil$.

FLYCOO maps each nonzero tensor element to a shard in each mode. A tensor of size $|T|$ with $N$ modes in the FLYCOO format is a sequence $x_0, \ldots, x_{|T|-1}$, where each element $x_i$ is a tuple $\langle s_i, p_i, val_i \rangle$, $s_i = (b_0, \ldots, b_{N-1})$ is a shard ID vector where each shard ID corresponds to a mode of the tensor. Here, $b_n = (j,q)$ if and only if $x_i \in shard_{n,j,q}$. 
This is used to locate the shards where each nonzero tensor element belongs in each mode. $p_i = (c_0, \ldots, c_{N-1})$ is the original indices of the nonzero tensor element in each dimension. $val_i$ is the value of the nonzero tensor elements of the tensor at $p_i$. Following the notation used in Section \ref{datamapping}, a single nonzero element in the FLYCOO format requires approximately $N \times \log_2 \left( \frac{|T|}{g} \right) + \sum_{h=0}^{N-1} \log_2 |I_h| + \beta_{\text{float}}$ bits, where $\beta_{\text{float}}$ is the number of bits needed to store the floating-point value of the nonzero tensor element. Here, $|s_i| \approx N \times \log_2 \left( \frac{|T|}{g} \right)$,  $|p_i| = \sum_{h=0}^{N-1} \log_2 |I_h|$, and $|val_i| = \beta_{\text{float}}$.
\vspace{-2mm}
\subsection{Dynamic Tensor Remapping} \label{datamapping}


During preprocessing, for each mode, nonzero elements of the tensor are assigned to a shard following Section~\ref{Tensor_Format_Definition}. When performing spMTTKRP mode by mode, the tensor elements are dynamically remapped based on the shard IDs associated with the mode to be executed next. 
Initially, the tensor is ordered according to the shard IDs of mode 0. During the spMTTKRP computation for mode 0, the tensor is reordered based on the shard IDs of mode 1. Therefore, by the time the computation for mode 1 begins, the tensor is already ordered according to mode 1. This reordering holds true for any mode during the computation process. As proven in~\cite{10.1145/3543622.3573179}, the dynamic tensor remapping requires $2 \times |T|$ external memory. This eliminates the need to create additional tensor copies equal to the number of tensor modes, which was previously necessary to facilitate mode-specific optimizations.
\vspace{-5mm}
\subsection{FLYCOO for multi-core CPU}\label{parameter_cpu}
We adapt the FLYCOO tensor format without modifying the format. We introduce several key contributions. (1) We propose Dynasor, a novel thread-level parallel algorithm that supports multi-core CPU-based spMTTKRP. This algorithm leverages super-shard-wise partition distribution among threads, ensuring load balancing across the workload. (2) We demonstrate that dynamic tensor remapping can be efficiently performed on a CPU-based hierarchical cache memory system, eliminating the need for a specialized memory system as proposed in~\cite{10.1145/3543622.3573179}. (3) We enable elementwise computation with dynamic tensor remapping in a single thread, with maximum parallelization possible on a CPU. (4) We show that FLYCOO can be adapted to general CPU platforms without specific hardware, like custom memory controllers.

In our work, we optimize the tensor partitioning parameters of the FLYCOO format for a given CPU platform. Consider a multi-core CPU with $\nu$ threads and a total cache size of $\Gamma$: Our goal is to select the tensor partitioning parameters of FLYCOO to (1) utilize all the threads optimally while executing Dynasor and (2) optimally share the cache among inputs and outputs of Dynasor. It can be achieved by satisfying the constraints shown in Equations~\ref{eqn_constraints_1} and~\ref{eqn_constraints_2}. Equations~\ref{eqn_constraints_1} and~\ref{eqn_constraints_2} follow the same notations as Section~\ref{Tensor_Format_Definition}.

\vspace{-5mm}
\begin{equation} \label{eqn_constraints_1}
\begin{split}
\forall n; \textsc{   } \frac{|I_n|}{m_n}= q \times \nu; q \in \mathbb{Z}^{+}
\end{split}
\end{equation}
\begin{equation} \label{eqn_constraints_2}
\begin{split}
\forall n; \textsc{   } \Gamma = \theta \times ((\alpha \times m_n \times R + \beta \times g) \times \nu + \sigma \times \sum_{j=0}^{k_n-1} \left\lceil\frac{|\textit{SS}_{n,j}|}{g}\right\rceil \\
0 < \theta < 1
\end{split}
\vspace{-5mm}
\end{equation}

The objective is to determine the optimal values for the variables $g$ and $m_n$ for each mode $n$ to efficiently utilize CPU threads and the CPU cache. All of these variables have inter-dependencies. For example, if $m_n$ of mode $n$ is too small, we are not able to choose larger $g$ values for very sparse tensors as the number of nonzero tensor elements of super-shards of mode $n$ can be significantly small. 

Equation~\ref{eqn_constraints_1} ensures that the number of super-shards is sufficiently large enough to be distributed among the available CPU threads for all the modes ($\forall n$). Equation~\ref{eqn_constraints_2} describes sharing the cache among different inputs to the proposed algorithm, Dynasor. $(\alpha \times m_n + \beta \times g) \times \nu$ guarantees the intervals that correspond to the super-shards currently being executed on the CPU threads at a given time and the input tensor elements correspond to the super-shards executing on the CPU threads fit inside the total CPU cache. Meanwhile, $\sigma \times \sum_{j=0}^{k_n-1} \left\lceil\frac{|\textit{SS}_{n,j}|}{g}\right\rceil$ make sure the memory address pointers in dynamic tensor remapping fit inside the total CPU cache. Here, $\alpha$, $\beta$, and $\sigma$ represent the size of a factor matrix row, nonzero tensor element, and address pointer for dynamic tensor remapping, respectively. We introduce $\theta$ ($ < 1$) to share the cache with input factor matrices. In our experiments, we set $\theta = 0.5$. Finally, we select a set of tensor partitioning parameters for a given input tensor that satisfies these requirements.
\section{Parallel Algorithm}\label{secparallel_algo}



\begin{algorithm}[ht]
    \DontPrintSemicolon
    Input: Input tensor ordered according to mode 0 shards, $\mathcal{H}_{0} = \{S_{0,j}:\forall j\}$ \;
    Super-shard to CPU thread map, $\{\mathcal{SS}\_List_{n}:\forall n\}$\;
    Randomly initialized factor matrices $\textbf{Y} = \{Y_0, Y_1,...,Y_{N-1}\}$\;
    Output: Updated factor matrices-set $\hat{\textbf{Y}} = \{\hat{Y}_0, \hat{Y}_1,...,\hat{Y}_{N-1}\}$ \;
    
    \For{each mode $n = 0, \ldots, N-1$} {
    
        Initialize $\hat{Y}_n$ as a zero matrix \;
        
        \For{each $S\_Map_{n,j}$ in $\mathcal{SS}\_List_{n}$ \textbf{parallel} } {
        
            \If{ $\text{thread}_{j}$ is idle}{

                \textcolor{blue}{// $len(\mathcal{SS}\_List_{n})$ $\leq$ number of threads} \;
            
                \For{each $S\_pointer \in S\_Map_{n,j}$}{
                
                    $S$ $\leftarrow$ \textbf{Load}($S\_pointer$) \;
                
                    \For{each element, $x_i = \langle s_i, p_i, val_i \rangle$ $\in S$}{
                    
                        \tcp{\textbf{-MTTKRP Computation-}} \;  
                        $value \leftarrow val_i$ \;
                        
                        $p_i = (c_0, \ldots, c_{N-1})$ \;
                        
                        $shard$\_$ids$, $s_i =$ ($b_0$, \ldots, $b_{N-1}$) \;
                        
                        $z \gets b_{y}$ \;
                        \textcolor{blue}{// $\ell$ is a vector of size \textit{R}} \;
                        $\ell \leftarrow \{1\}$ \;
                        \For{ input mode $w\in\{0,\ldots,N-1\}\setminus\{n\}$}{
                        
                            $vec \leftarrow $ \textbf{Load}(row $c_w$ from $w^\text{th}$ factor matrix) \;
                            \For{each rank $r$ in $R$ \textbf{parallel}}{
                                $\ell(r) \leftarrow \ell(r) \times vec(r)$ \; 
                            }
                        }
                        \For{each rank $r$ in $R$ \textbf{parallel}}{
                            $\hat{Y}_n(c_n, r) \leftarrow \hat{Y}_n(c_n, r) + value \times \ell(r)$ \;
                        }
                        \tcp{\textbf{-Dynamic Tensor Remapping-}} \;
                         \textbf{Store}($x_i$ at shard$_{b_{(n+1) mod N}}$) \;
                        
                    }
                    
                }
            }
        }

    }
\caption{Dynasor: Algorithm for spMTTKRP with \underline{Dyna}mic Ten\underline{sor} Remapping on multi-core CPU}
\label{parallel_alg}
\end{algorithm}

In Algorithm~\ref{parallel_alg}, tensor ordered according to mode 0 shards ($\mathcal{H}_{0}$), factor matrices ($\textbf{Y} = \{Y_0, Y_1,...,Y_{N-1}\}$), and super-shard to CPU thread map ($\mathcal{SS}\_List$) are used as the inputs. The preprocessing involves converting the input tensor to FLYCOO format, during which the super-shard-to-shard mapping for each mode is generated as metadata. This mapping is essentially a dictionary that indicates which shard belongs to which super-shard. $\mathcal{SS}\_List$ is used as a static scheduling policy of the super-shards among the available threads. More comprehensive details regarding the scheduling process can be found in Section~\ref{load_balancing}.

In Algorithm~\ref{parallel_alg}, a CPU with $\nu$ threads can simultaneously process $\nu$ super-shards. All threads are synchronized after each mode. Once a super-shard is allocated to a CPU thread, each thread undertakes two distinct operations on every nonzero tensor element within the super-shard. These operations consist of the following:
(1) elementwise spMTTKRP computation (see Section~\ref{sec_elementwise_compute}), and
(2) dynamic tensor remapping.

Using $\mathcal{SS}\_List$, each thread sequentially loads the shards of the assigned super-shards into the CPU cache (Algorithm~\ref{parallel_alg}, line 11). Firstly, the index values of all the modes are extracted from the tensor element (Algorithm~\ref{parallel_alg}, line 15). Next, the corresponding rows of the input factor matrices are loaded from the external memory (Algorithm~\ref{parallel_alg}, lines 20-21). Subsequently, the current CPU thread performs elementwise operations between the tensor element and each element of the rows of the input factor matrices (Algorithm~\ref{parallel_alg}, lines 22-25). The elementwise computation can be summarized as Equation~\ref{elementwise_eqn}. Equation~\ref{elementwise_eqn} uses the same notation as the Algorithm~\ref{parallel_alg}. Here, elementwise computation is performed for the $r^{\text{th}}$ column element of $c_n^{th}$ row of $Y_{n}$ factor matrix. 
\vspace{-3mm}
\begin{equation}
\hat{Y}_{n,c_n,r}= \hat{Y}_{n,c_n,r} + value \times  \prod_{w=0;  w \neq n}^{N-1} \hat{Y}_{w,c_w,r};
0 \leq r < R
\label{elementwise_eqn}
\vspace{-1mm}
\end{equation}

After computing spMTTKRP for mode $n$, the CPU executes spMTTKRP for the subsequent mode, which is given by $(n+1)$ mod $N$. To support the proposed parallel algorithm, nonzero tensor elements in the input tensor must be ordered according to the output mode shard ids. Therefore, the tensor is remapped based on the shard IDs of mode $(n+1) \mod N$ to compute the spMTTKRP for the upcoming mode. As a result, while the current mode is being executed, the tensor is remapped in parallel according to the shard IDs of the upcoming mode, facilitating the sequential execution of all the modes (Algorithm~\ref{parallel_alg}, line 27). The algorithm maintains a record of the locations that need to be filled for each shard of the upcoming mode, determining the memory location to be filled in each shard and allowing for dynamic tensor remapping.

By following this process, the algorithm leverages the parallelism offered by the multiple threads to process multiple super-shards concurrently while ensuring synchronization and orderly execution of the required operations.

Initially, the input tensor is arranged in the CPU external memory based on the shard IDs of mode 0, denoted as $\mathcal{H}_{0} = \{SS_{0,j}: \forall j\}$. Meta-data is used as inputs to establish the mapping between threads and shards, enabling efficient computation scheduling across multiple threads. This mapping is represented as ${\mathcal{SS}\_List_{n,j} : \forall n,j}$. It is worth noting that both $\mathcal{SS}\_List_{n}$ and $S\_Map_{n,j}$ (Algorithm~\ref{parallel_alg}, line 2 and line 10) are arrays are pointers that reference to tensor partitions.

Algorithm~\ref{parallel_alg} is designed for a multi-thread CPU, where each thread can independently perform the elementwise computations on each assigned super-shard without interfering with the other threads (i.e., lock-free computation). Lock-free computation is enabled by collecting all the nonzero tensor elements of an output mode index to the same super-shard. It ensures that all the elementwise computations used to update a row of an output factor matrix are executed on the same CPU thread.

\vspace{-2mm}
\subsection{Super-shard Scheduling Scheme}\label{load_balancing} \label{shard_scheduling}

We propose static super-shard scheduling to ensure balanced workload distribution among the CPU threads. During the preprocessing time, we map each super-shard and its shards to a CPU thread and generate $\mathcal{SS}\_List$. $\mathcal{SS}\_List$ is a 2-dimensional list that indicates the super-shards assigned to each CPU thread in each mode. We employ Algorithm~\ref{ss_scheduling}, a greedy algorithm to distribute the super-shards among the CPU threads, which achieves balanced workload distribution for each mode. By distributing the super-shards appropriately, the proposed method aims to optimize the utilization of CPU resources and achieve efficient parallel execution of the spMTTKRP computation for each mode.

\begin{algorithm}[ht]
\DontPrintSemicolon
Input: A sorted list of super-shards, \textit{SS} depending on the number of shards in the super-shard. Here, $\textit{SS}_{n,j}$ indicates the $j^{th}$ super-shard in mode $n$. $|\textit{SS}_{n,j}|$ indicates the number of shards belongs to the super-shard $\textit{SS}_{n,j}$\;
$\mathcal{SS}\_List$ with $N \times \nu$ empty bins. Here, $\mathcal{SS}\_List_{n,j}$ indicates the $j^{th}$ bin of mode $n$. $j^{th}$ bin correspond to the $j^{th}$ CPU thread\;
Output: $\mathcal{SS}\_List$,  where each super-shards, \textit{SS} mapped to a bin inside $\mathcal{SS}\_List$\;

\For{each mode $n = 0, \ldots, N-1$} {
    \For{each super-shard id $j = 0, \ldots, k_{n-1}$} {
        \textcolor{blue}{// identify the least filled bin in mode $n$ of $\mathcal{SS}\_List$ } \;
        $ind =$ Index$(min(|\mathcal{SS}\_List_{n,i}|))$; $\forall i$ \;
        $\mathcal{SS}\_List_{n,ind}.append(\mathcal{SS}_{n,j})$ \;
    }
}
\Return $\mathcal{SS}\_List$
\caption{Super-shard scheduling among threads}
\label{ss_scheduling}
\end{algorithm}

In this context, the total number of computations associated with a super-shard is directly proportional to the number of nonzero tensor elements it contains. Since each super-shard is partitioned into shards with an equal number of tensor elements, the total number of computations is also proportional to the number of shards within a super-shard. It is worth noting that the number of shards in a super-shard can vary based on the sparsity of the tensor.

Suppose a tensor of size $T$ is partitioned into super-shards $\{SS_{n,j} : \forall \, j\}$ for a mode $n$. Let $|SS_{n,j}|$ be the number of shards in the super-shard $SS_{n,j}$. We reorder the indices of super-shards such that $|SS_{n, j}| \ge |SS_{n,j'}|$ if $j < j'$ so that they are sorted in descending order of the number of shards. We use the sorted list $SS$ as the input to the Algorithm~\ref{ss_scheduling}. Each super-shard ($SS_{n,j}$) is iteratively assigned to the CPU thread currently the least heavily loaded (i.e., with the least number of shards assigned). We perform the above operation for all the tensor modes as indicated in Algorithm~\ref{ss_scheduling}. For a given output mode, let $|\overline{SS}|_{\textrm{max}}$ be the largest number of shards assigned to a single thread, and $|\overline{SS}|^*_{\textrm{max}}$ be the value of $|\overline{SS}|_{\textrm{max}}$ in an optimal shard distribution among the threads. Then our proposed greedy approach above guarantees that $|\overline{SS}|_{\textrm{max}} \le 4/3 \cdot |\overline{SS}|^*_{\textrm{max}}$~\cite{graham1969bounds}.

\vspace{-3mm}
\subsection{Memory Requirements}

In this Section, we use the same notations as Section~\ref{Data_partitioning}. Dynasor requires  $2 \times |T|$ to store the tensor, including the additional memory required to perform dynamic tensor remapping. Dynasor also keeps the total elements of factor matrices equal to $|\text{\textbf{I}}| \times R$, where $|\text{\textbf{I}}| = \sum_{n=0}^{N-1} |I_n|$. As the number of super-shards is constructed to be the same as the number of intervals (see Section~\ref{Tensor_Format_Definition}), the total number of super-shards equal to $\frac{|I_n|}{m_n}$ for mode $n$. Since $\mathcal{SS}\_List_{n}$ used for scheduling (see Section~\ref{shard_scheduling}) requires keeping pointers for each super-shard, it requires $\sum_{n=0}^{N-1} \frac{|I_n|}{m_n} $  pointers. Dynamic tensor remapping required pointers for each shard (Algorithm~\ref{parallel_alg}, line 27) resulting in a total of $\sum_{j=0}^{k_n-1} \left\lceil|\textit{SS}_{n,j}|/g\right\rceil$ shard pointers.

\section{Experimental Results}
\vspace{-1mm}

\subsection{Experimental Setup}
\subsubsection{Platforms}\label{sec_platform}
We conduct detailed experiments on Intel Xeon Gold 5120 CPU with Skylake microarchitecture. The platform consists of two sockets, each CPU consisting of 14 physical cores (28 threads) running at a fixed frequency of 2.2 GHz, sharing 128 GB of CPU external memory. Utilizing this platform, we systematically vary hardware parameters such as the total CPU external memory and the number of CPU threads. To quantify the improvements, we measure CPU thread utilization and CPU external memory utilization.

We also demonstrate superior performance in total execution time on AMD platforms. We use AMD Ryzen Threadripper 3990X CPU with Zen 2 microarchitecture. The platform consists of two sockets, each CPU consisting of 32 physical cores (64 threads) running at a fixed frequency of 2.2 GHz, sharing 128 GB of CPU external memory.

Both systems run Ubuntu 18.04.5 LTS Linux distribution as the Operating System.


\subsubsection{Implementation}
The code is built using g++ 7.5.0 C/C++ compiler (version 19.1.3) and OpenMP application programming interface. The experiments use all hardware threads on the target platforms unless otherwise stated. 

We use the Linux perf~\cite{Melo2010TheNL} and Intel Advisor~\cite{o2017intel} for performance counter measurements, thread pinning, and roofline analysis on the Intel platform.

\subsubsection{Datasets}
Similar to state-of-the-art~\cite{10.1145/3543622.3573179, alto_paper, 8665782}, we use tensors from the Formidable Repository of Open Sparse Tensors and Tools (FROSTT) dataset~\cite{frosttdataset}. Table~\ref{table3} shows a summary of the characteristics of the tensors.

 \begin{table}[ht]
 \vspace{-2mm}
\caption{Characteristics of the sparse tensors}
\vspace{-4mm}
\begin{center}
\resizebox{\columnwidth}{!}{
\begingroup
\setlength{\tabcolsep}{6pt} 
\renewcommand{\arraystretch}{1.1} 
\begin{tabular}{ |c|c|c|c|c| }
 \hline
 \textbf{Tensor} & \textbf{Shape} & \#\textbf{NNZs} & \textbf{Density} \\
 \hline\hline
 Nell-1 & $2.9M \times 2.1M \times 25.5M$ & $143.6M$ & $9.1 \times 10^{-13}$ \\ 
 \hline
 Nell-2 & $12.1K \times 9.2K \times 28.8K$ & $76.9M$ & $2.4 \times 10^{-05}$ \\ 
 \hline
  Flickr & $319.6K \times 28.2M \times 1.6M$ & $112.9M$  & $1.1 \times 10^{-14}$ \\
 \hline
Delicious & $532.9K \times 17.3M \times 2.5M \times 1.4K$ & $140.1M$  & $4.3 \times 10^{-15}$ \\
 \hline
 Vast & $165.4K \times 11.4K \times 2 \times 100 \times 89 $ & $26M$  & $7.8 \times 10^{-07}$ \\
 \hline
\end{tabular}
\endgroup
}
\label{table3}
\end{center}
\vspace{-5mm}
\end{table}

\subsubsection{Baselines}
We evaluate our approach against state-of-the-art CPU-based work ALTO~\cite{alto_paper}, HiCOO~\cite{8665782}, and STeF~\cite{9820702}. ALTO, HiCOO, and Dynasor (i.e., our work) are executed on an identical CPU environment. Additionally, we compare against the reported results of STeF in~\cite{9820702}.

To obtain the best results with HiCOO, we follow the recommended configurations of the source code~\cite{hicoo_repo}. However, the HiCOO code~\cite{hicoo_repo} only supports tensors with up to 4 modes. Therefore, we use a benchmarking tool~\cite{li2020pasta} provided by the authors of HiCOO~\cite{8665782} to estimate the execution time for tensors with more than 4 modes. In our experiments, we use open-source ALTO repository~\cite{alto_github} compiled with GCC and BLAS library~\cite{blackford2002updated}.

\subsubsection{Tensor Partitioning Parameters of FLYCOO}

When $|I_n|$ is less than $\nu$, we set $m_n$ equal to 1. Conversely, when $|I_n|$ is much larger than $\nu$, we determine the value of $q$ from Equation~\ref{eqn_constraints_1}, such that $1000 < m_n < 16000$ holds true for both CPU platforms. Additionally, we ensure that the value of $g$ falls within the range of 1024 to 32768 while satisfying Equation~\ref{eqn_constraints_2} for all datasets on both CPU platforms.

\subsection{Implementing Dynamic Remapping}

\begin{wrapfigure}{r}{0.27\textwidth}
\vspace{-5mm}
  \begin{center}
    \includegraphics[width=0.27\textwidth]{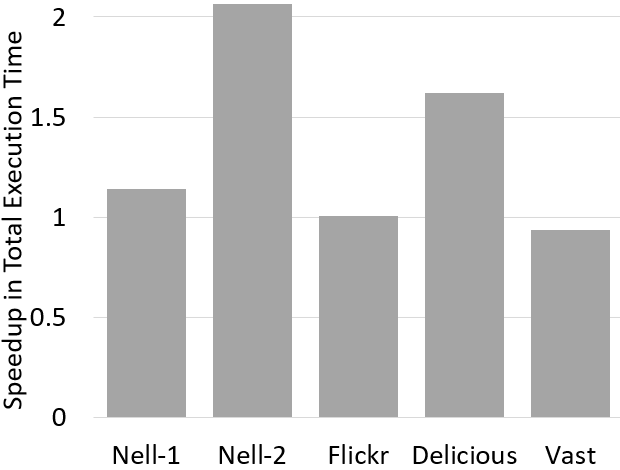}
  \end{center}
  \vspace{-4mm}
  \caption{Impact of performing dynamic tensor remapping and elementwise spMTTKRP computation in the same CPU thread vs. different CPU threads}
  \label{thread-share}
  \vspace{-3mm}
\end{wrapfigure}

In Algorithm~\ref{parallel_alg}, we can perform the elementwise computation and dynamic tensor remapping integrated into the same thread or independently in different threads, overlapping the execution of elementwise computation and dynamic tensor remapping. Figure~\ref{thread-share} illustrates the overall improvement in execution time achieved by performing both operations on the same thread compared to running them on separate threads. Throughout the paper, we use integrated remapping and elementwise computation as it shows better speedups.

\begin{figure*}

\subfigure{\hspace{5cm} \includegraphics[width=0.4\textwidth]{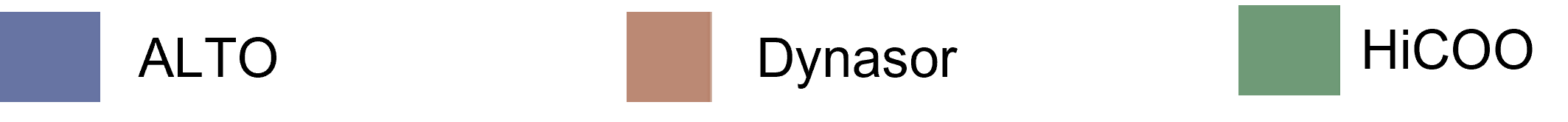}}

\setcounter{subfigure}{0}
\begin{adjustbox}{varwidth=\textwidth,fbox,center}
    \subfigure[Nell-1]{\includegraphics[width=0.20\textwidth]{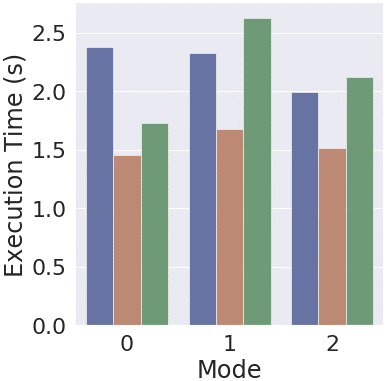}} 
    \rulesep
    \subfigure[Nell-2]{\includegraphics[width=0.18\textwidth]{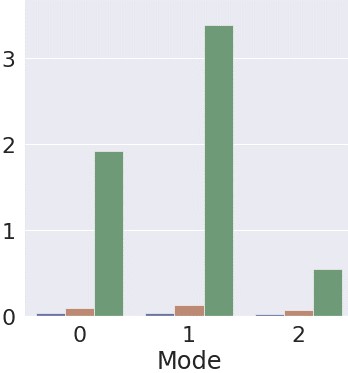}} 
    \rulesep
    \subfigure[Flickr]{\includegraphics[width=0.18\textwidth]{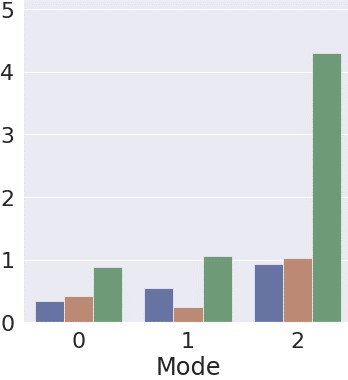}} 
    \rulesep
    \subfigure[Delicious]{\includegraphics[width=0.18\textwidth]{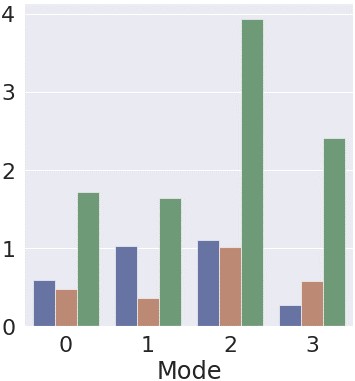}}
    \rulesep
    \subfigure[Vast]{\includegraphics[width=0.18\textwidth]{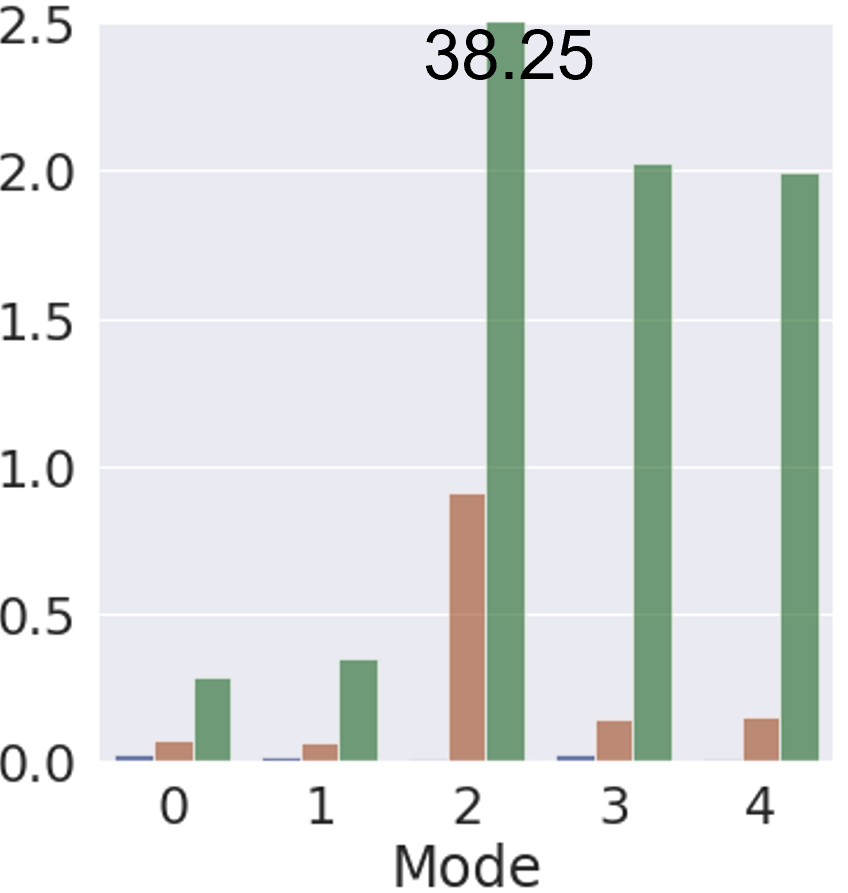}} 
    \centering \textsc{ $\textit{\textbf{R = 16}}$ }
\end{adjustbox}

\begin{adjustbox}{varwidth=\textwidth,fbox,center}
    \subfigure{\includegraphics[width=0.20\textwidth]{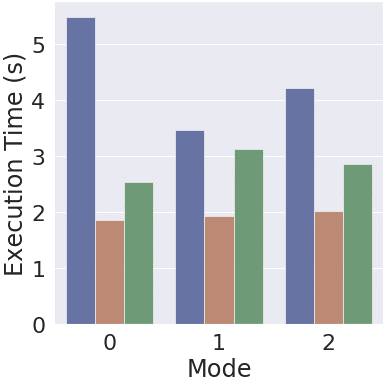}} 
    \rulesep
    \subfigure{\includegraphics[width=0.18\textwidth]{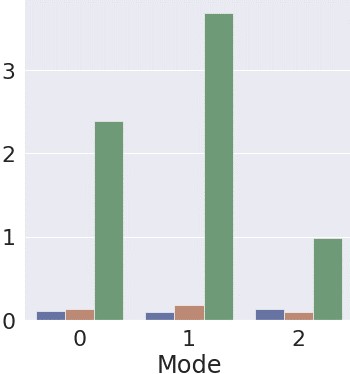}} 
    \rulesep
    \subfigure{\includegraphics[width=0.18\textwidth]{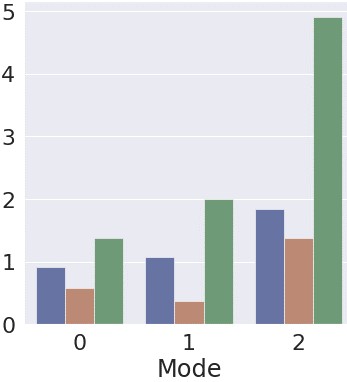}} 
    \rulesep
    \subfigure{\includegraphics[width=0.18\textwidth]{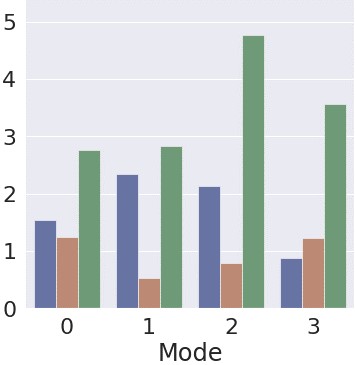}}
    \rulesep
    \subfigure{\includegraphics[width=0.18\textwidth]{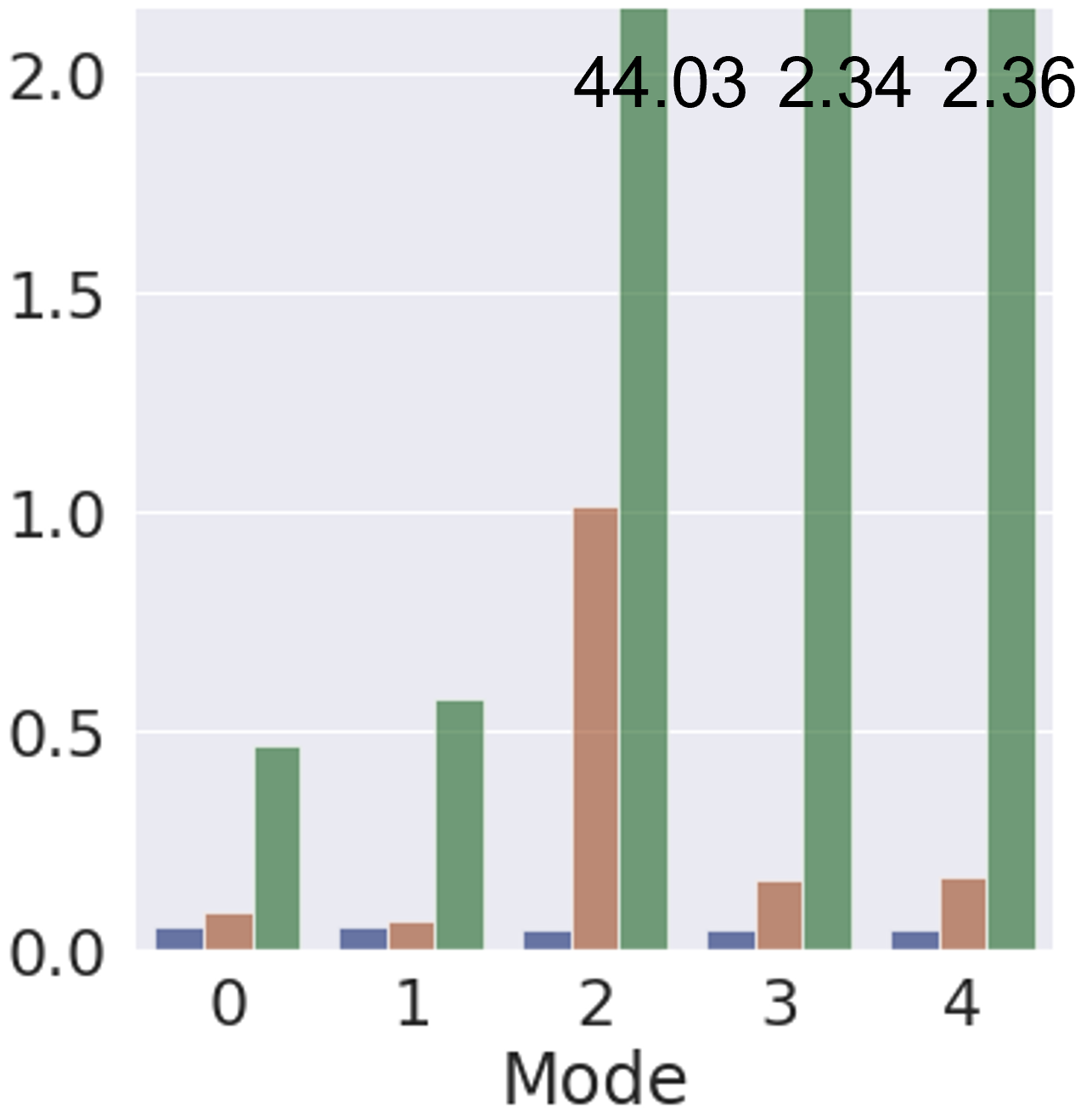}} 
    \centering \textsc{ $\textit{\textbf{R = 32}}$ }
\end{adjustbox}

\begin{adjustbox}{varwidth=\textwidth,fbox,center}
    \centering
    \subfigure{\includegraphics[width=0.20\textwidth]{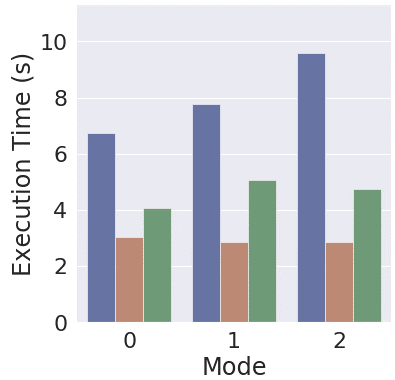}} 
    \rulesep
    \subfigure{\includegraphics[width=0.18\textwidth]{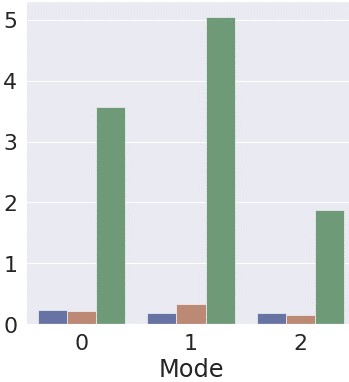}} 
    \rulesep
    \subfigure{\includegraphics[width=0.18\textwidth]{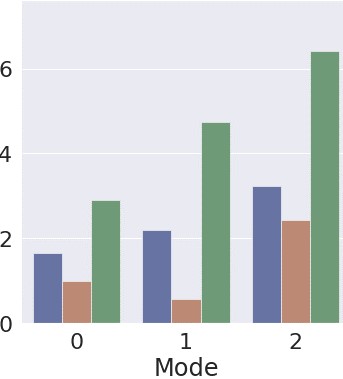}} 
    \rulesep
    \subfigure{\includegraphics[width=0.18\textwidth]{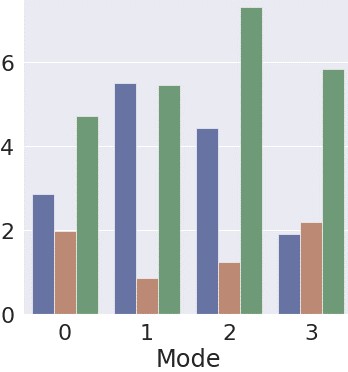}}
    \rulesep
    \subfigure{\includegraphics[width=0.18\textwidth]{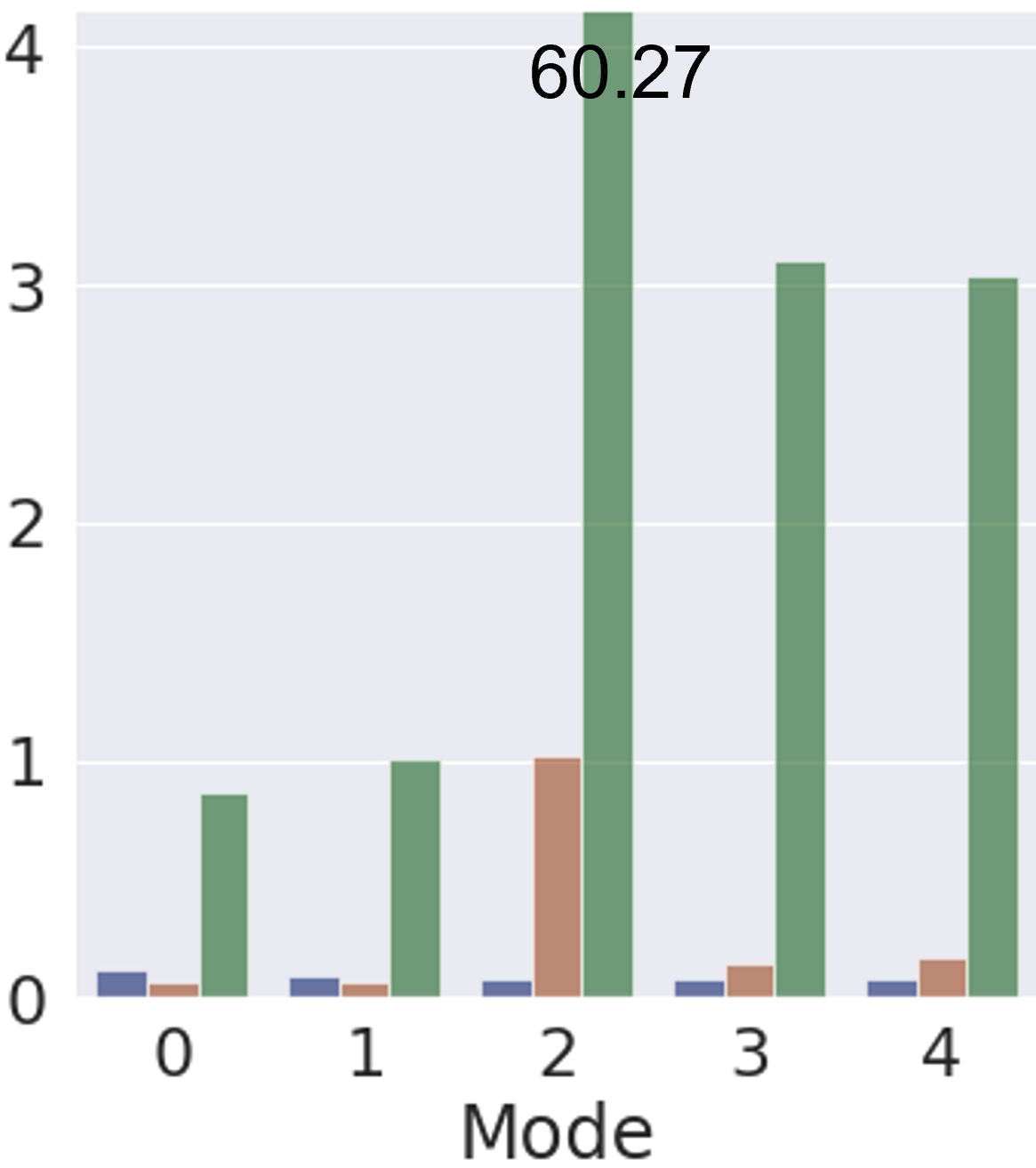}} 
    \centering \textsc{ $\textit{\textbf{R = 64}}$ }
\end{adjustbox}

\begin{adjustbox}{varwidth=\textwidth,fbox,center}
    \centering
    \subfigure{\includegraphics[width=0.20\textwidth]{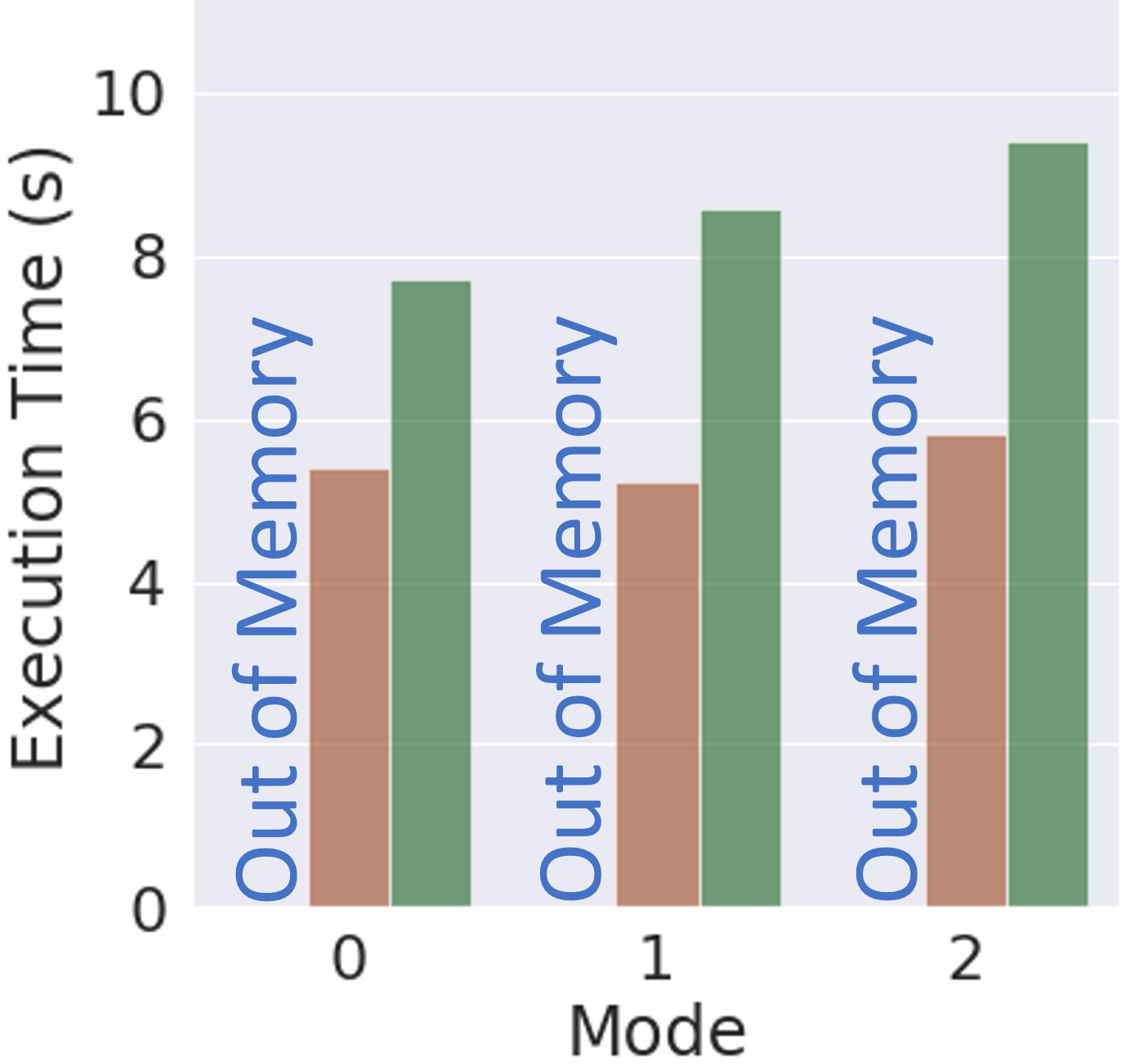}} 
    \rulesep
    \subfigure{\includegraphics[width=0.18\textwidth]{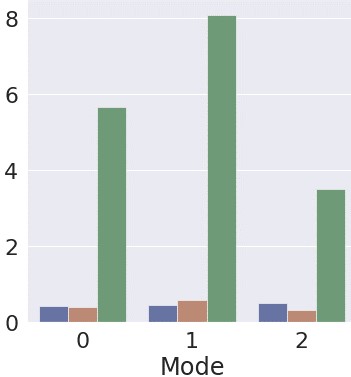}} 
    \rulesep
    \subfigure{\includegraphics[width=0.18\textwidth]{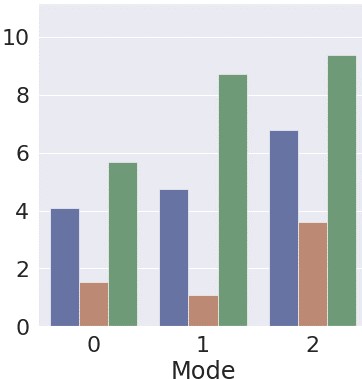}} 
    \rulesep
    \subfigure{\includegraphics[width=0.18\textwidth]{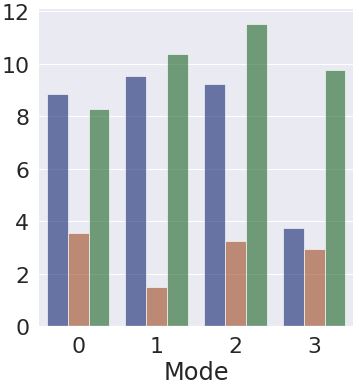}}
    \rulesep
    \subfigure{\includegraphics[width=0.18\textwidth]{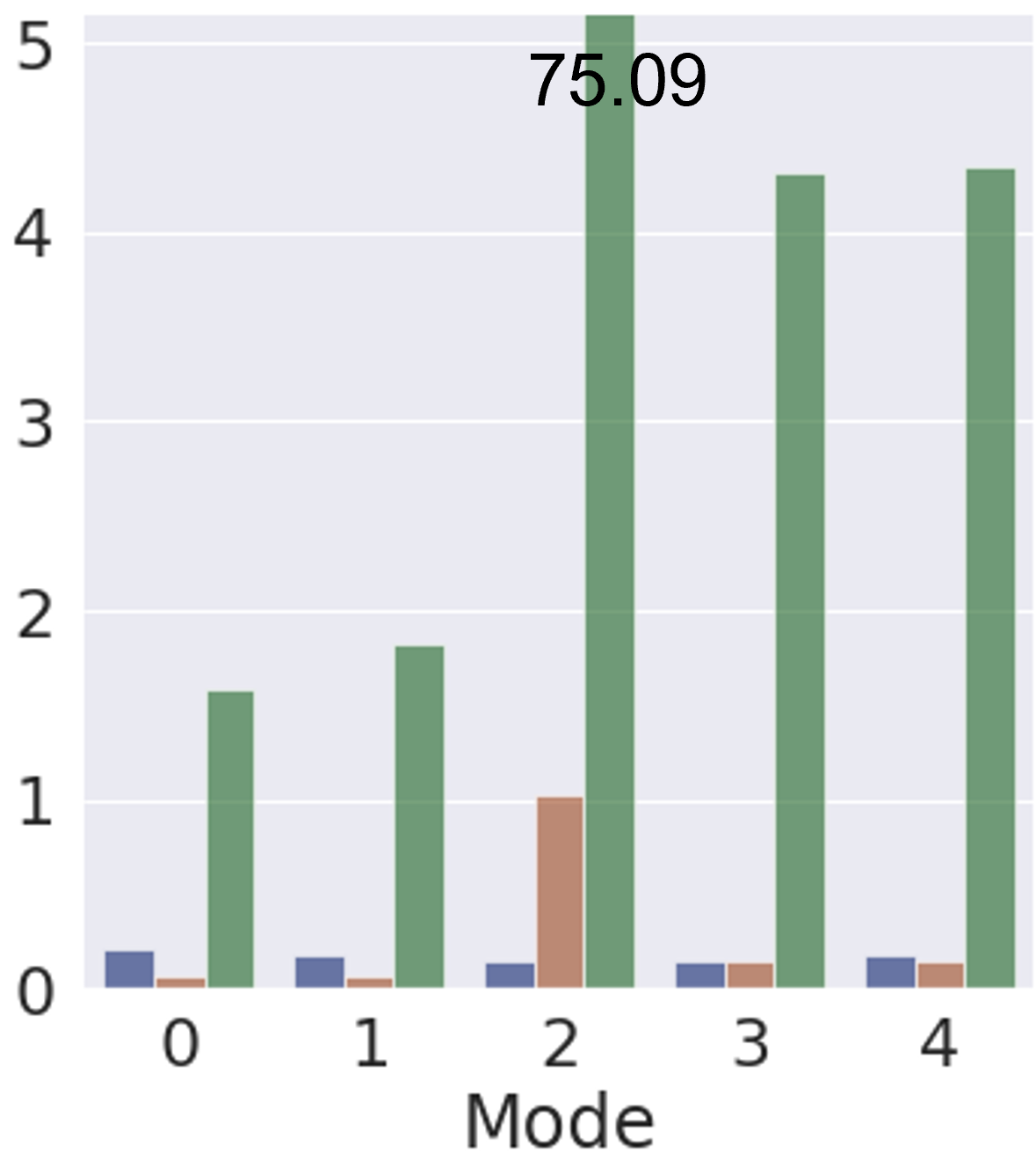}} 
    \centering \textsc{ $\textit{\textbf{R = 128}}$ }
\end{adjustbox}

\begin{adjustbox}{varwidth=\textwidth,fbox,center}
    \centering
    \subfigure{\includegraphics[width=0.20\textwidth]{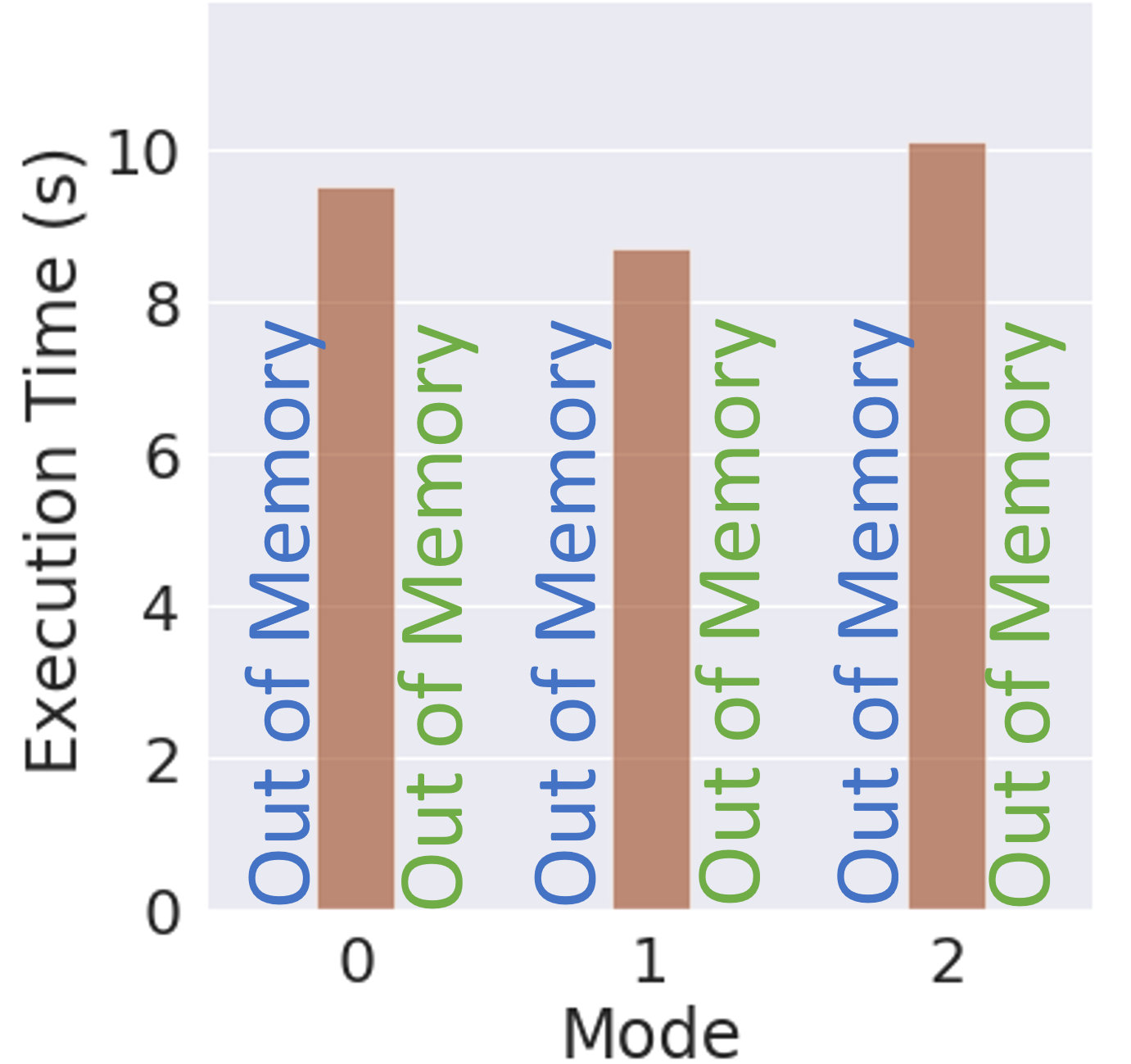}} 
    \rulesep
    \subfigure{\includegraphics[width=0.18\textwidth]{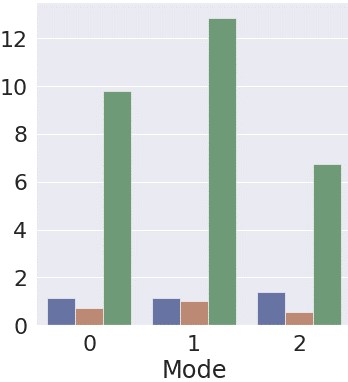}} 
    \rulesep
    \subfigure{\includegraphics[width=0.18\textwidth]{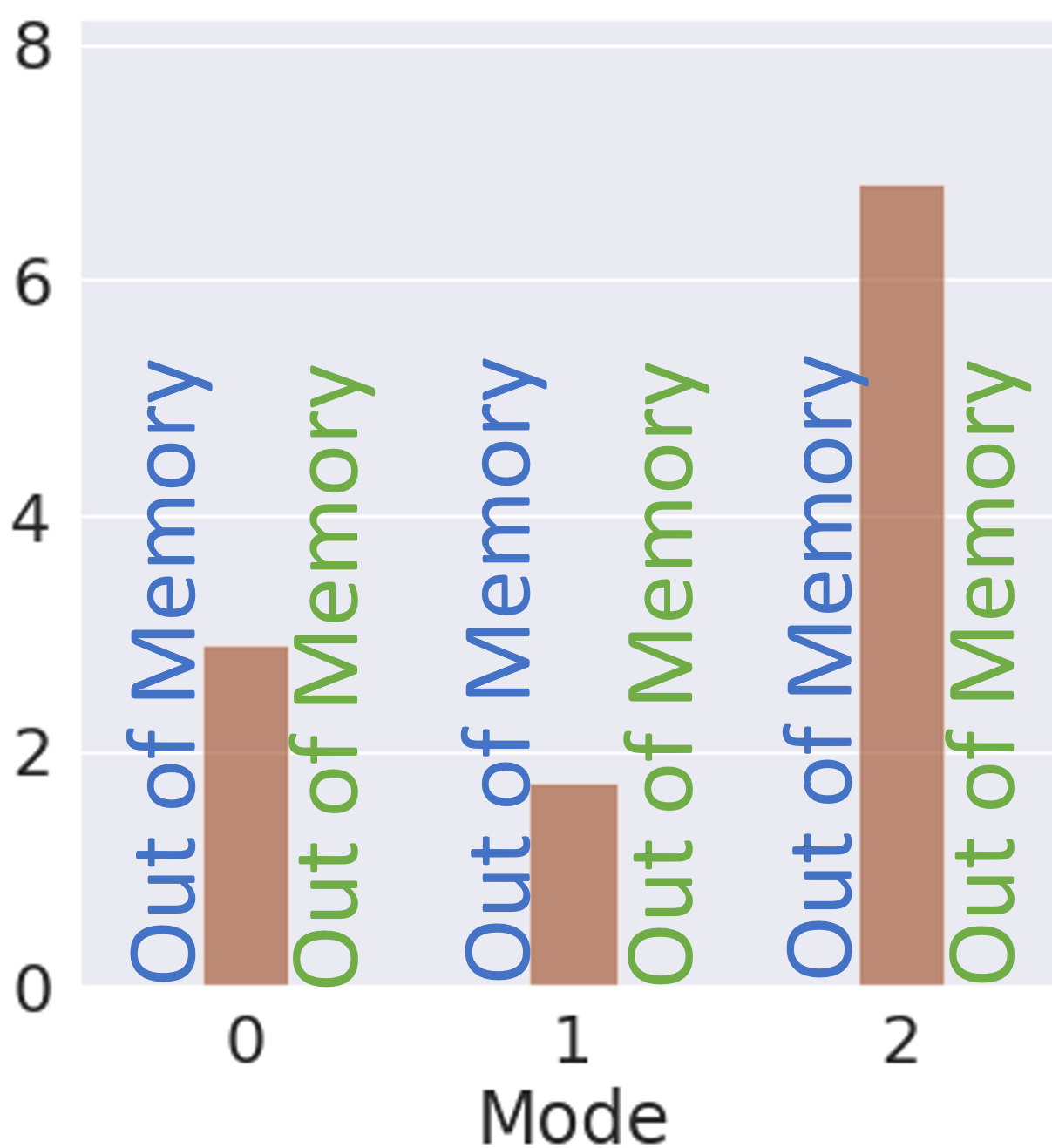}} 
    \rulesep
    \subfigure{\includegraphics[width=0.18\textwidth]{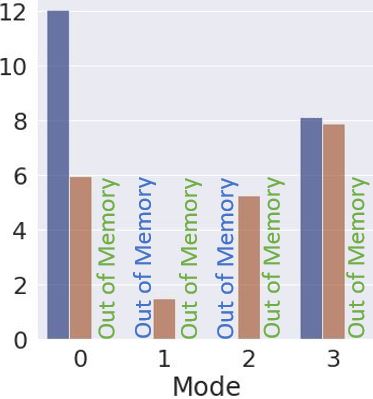}}
    \rulesep
    \subfigure{\includegraphics[width=0.18\textwidth]{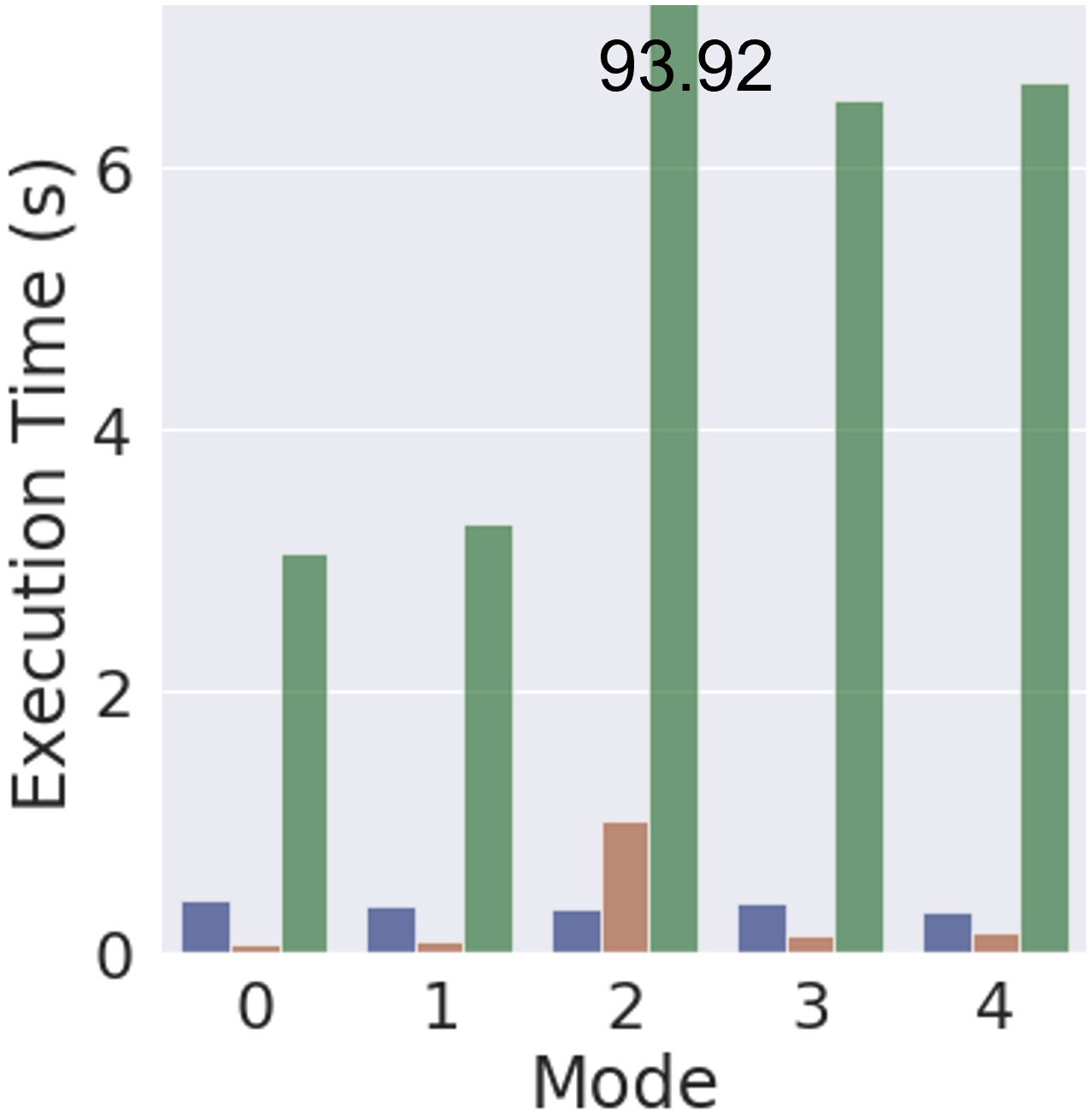}} 
    \centering \textsc{ $\textit{\textbf{R = 256}}$ }
\end{adjustbox}
	
    \caption{Total execution time on Intel platform}
    \label{dataset_256}
\end{figure*}

\begin{figure*}
    \centering
    \subfigure{\includegraphics[width=0.20\textwidth]{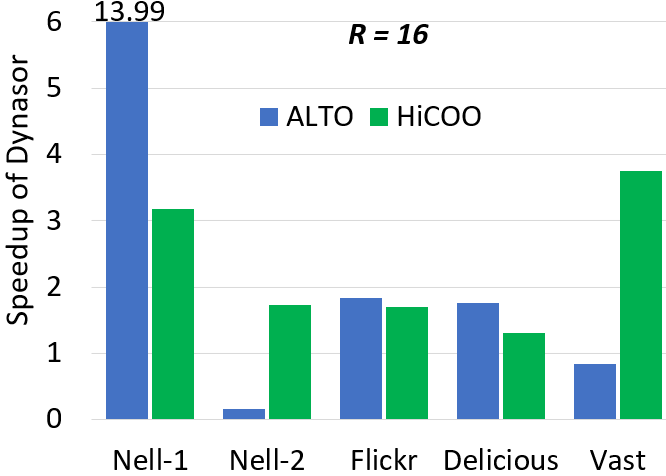}} 
    \subfigure{\includegraphics[width=0.18\textwidth]{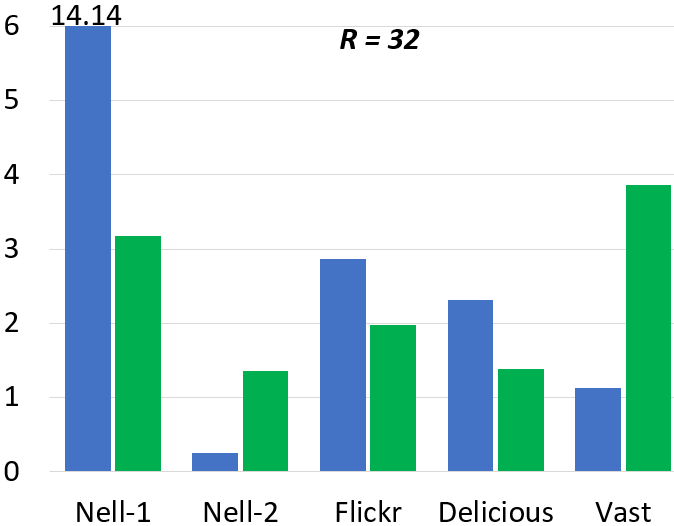}} 
    \subfigure{\includegraphics[width=0.18\textwidth]{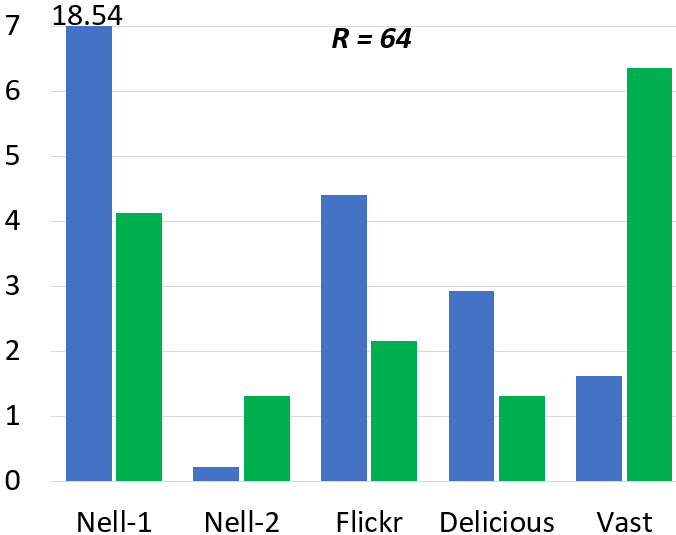}} 
    \subfigure{\includegraphics[width=0.18\textwidth]{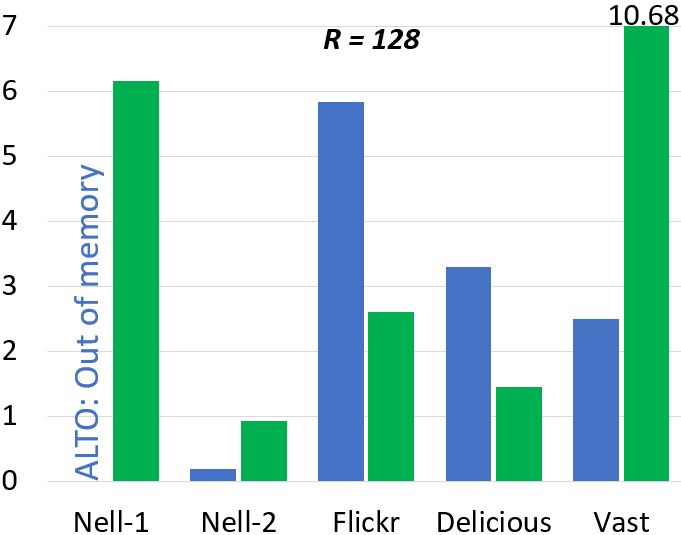}}
    \subfigure{\includegraphics[width=0.18\textwidth]{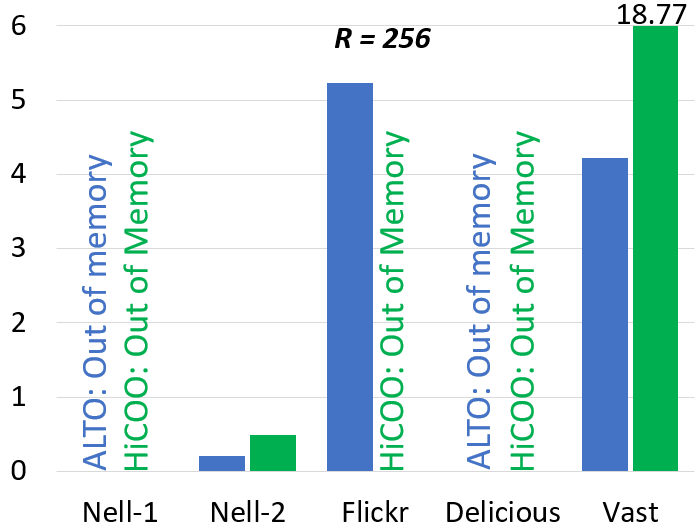}} 
    \vspace{-4mm}
    \caption{Speedup of Dynasor against state-of-the-art baselines on AMD platform}
    \label{amd_speedup}
\end{figure*}

\subsection{Overall Performance}

\begin{table*}
\vspace{-3mm}
  \caption{Speedup of Dynasor over state-of-the-art}
  \centering
  \begin{tabular}{|c|c|c|c|c|c|c|}
    \hline
      & $R = 16$ & $R = 32$ & $R = 64$ & $R = 128$ & $R = 256$ & Overall Speedup \\
    \hline
    Speedup over ALTO~\cite{alto_paper} on Intel platform & 0.88 & 1.47 & 1.92 & 2.25$^+$ & 2.36$^+$ & \textbf{1.70}$^+$ \\
    \hline
    Speedup over HiCOO~\cite{8665782} on Intel platform & 9.02 & 9.19  & 11.51 & 14.34 & 37.55$^*$ & \textbf{13.67}$^*$ \\
    \hline
    Speedup over ALTO~\cite{alto_paper} on AMD platform & 3.75 & 4.11 & 5.53 & 2.95$^+$ & 3.21$^+$ & \textbf{3.22}$^+$ \\
    \hline
    Speedup over HiCOO~\cite{8665782} on AMD platform & 2.33 & 2.35  & 3.03 & 4.37 & 9.63$^*$ & \textbf{4.34}$^*$ \\
    \hline
    Speedup over STeF~\cite{9820702} & n/a & 1.71  & 2.52 & n/a & n/a & \textbf{2.12} \\
    \hline
    Overall Speedup & \textbf{4.00} & \textbf{3.77} & \textbf{4.91} & \textbf{5.98}$^+$ & \textbf{13.19}$^{+*}$ & \textbf{6.37}$^{+*}$ \\
    \hline
  \end{tabular}

    \vspace{1mm}
    $+$: ALTO runs out of memory for some input tensors \\
    \hspace{0.5mm} $*$: HiCOO runs out of memory for some input tensors \\
    \hspace{-36mm} n/a: Not reported in~\cite{9820702} 
  \label{tab:speed-up}
\end{table*}



Figure~\ref{dataset_256} compares the total execution time of ALTO, HiCOO, and Dynasor for all the datasets on the Intel CPU. The factor matrix rank ($R$) in the experiments varies from 16 to 256.

\begin{figure}
\vspace{-5mm}
    \centering
    \subfigure{\includegraphics[width=0.23\textwidth]{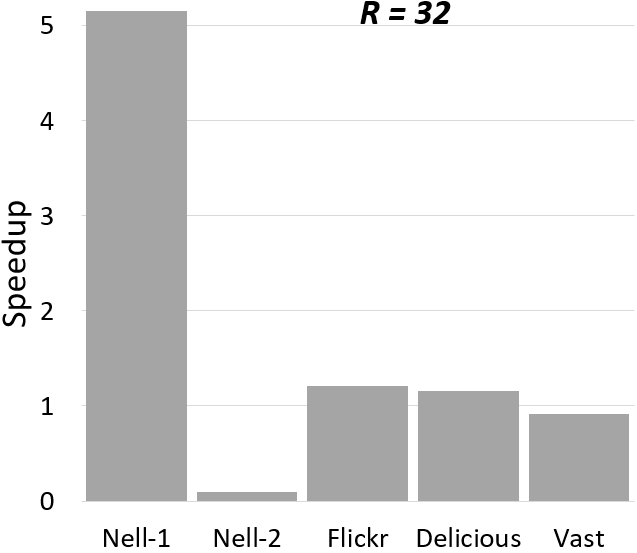}} 
    \subfigure{\includegraphics[width=0.22\textwidth]{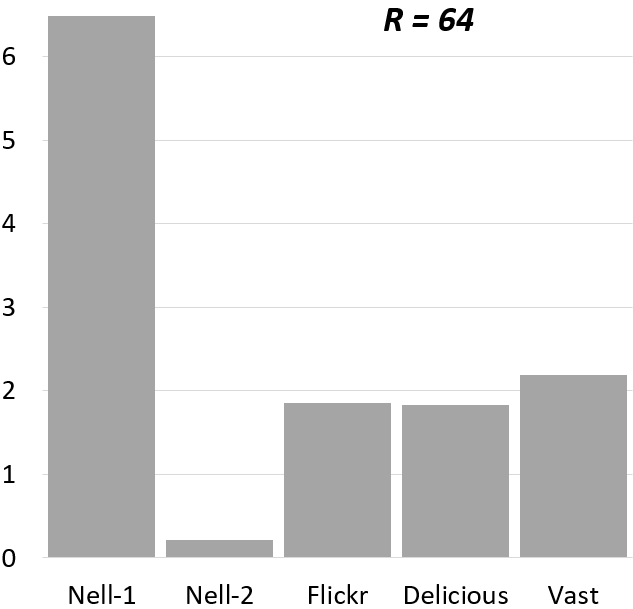}} 
    \vspace{-3mm}
    \caption{Speedup compared with STeF}
    \label{fig:stef_time}
    \vspace{-8mm}
\end{figure}

For Flickr and Nell-1, ALTO and HiCOO ran out-of-memory while computing factor matrices for larger ranks (e.g., $R = 256$) since the intermediate values generated during the computation exceed the available CPU external memory (128 GB). In contrast, Dynasor is able to compute the factor matrices for all the ranks. Super-shards-wise computation in Dynasor avoids such memory explosion as it allows intermediate values to be fit within the CPU cache during runtime.

Dynasor exhibits superior overall performance in total computation time, except for a few specific cases. For example, Dynasor partitions the indices in each mode equally among the super-shards, which are then distributed across the CPU cores for computation. In mode 3 of the Vast dataset, there are only 2 indices that limit the distribution of the workload across all the threads. As a result, ALTO outperforms our approach in mode 3 of the Vast dataset. In Nell-2, each mode contains only thousands of indices (i.e., $12092$, $9184$, and $28818$) which are smaller than other large datasets with millions of indices per mode. Consequently, all the factor matrices of Nell-2 fit inside the CPU cache. Therefore, combining intermediate values in Adaptive Linearized Order (in ALTO) becomes an inexpensive operation compared to dynamic remapping use in Dynasor. Therefore, ALTO outperforms Dynasor for Nell-2.

\begin{wrapfigure}{r}{0.22\textwidth}
  \begin{center}
    \includegraphics[width=0.22\textwidth]{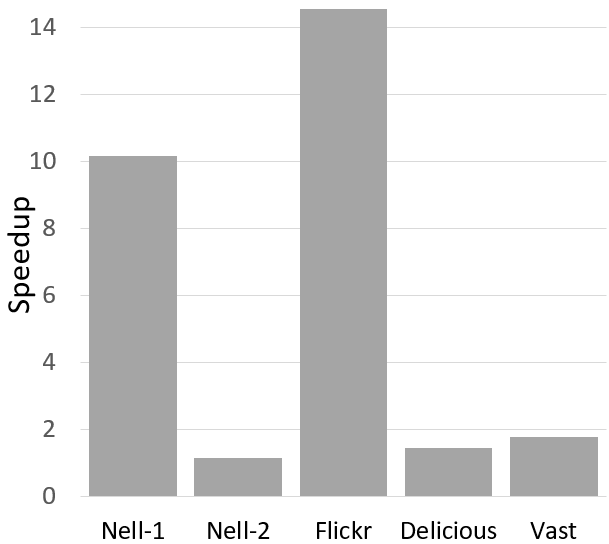}
  \end{center}
\vspace{-3mm}
  \caption{Impact of proposed scheduling strategy}
\vspace{-5mm}
\label{impact_ss_schedule}
\end{wrapfigure}

To demonstrate platform independence, we conducted identical experiments on an AMD Ryzen Threadripper 3990X CPU platform. As shown in Figure~\ref{amd_speedup}, the total execution time exhibits a similar trend as the Intel platform. The overall speedup of Dynasor is 3.22$\times$ and 4.32$\times$ compared with the baselines (HiCOO and ALTO) executed on the same AMD platform.

Based on the results reported in Kurt et al.~\cite{9820702}, we compare our results and STeF. We employ the identical AMD platform as in~\cite{9820702}. Figure~\ref{fig:stef_time} illustrates the results, showcasing that our approach surpasses STeF in terms of overall performance. 
STeF incorporates a selective intermediate value storage technique during computation, while our approach leverages a dynamic tensor remapping technique, resulting 2.12$\times$ average speedup in total execution time. In datasets like Nell-2, since there are only a few thousand indices along all the modes, the selective intermediate value storage technique used in STeF outperforms the dynamic tensor remapping technique used in Dynasor.

Compared with the baselines, our approach achieves an average speedup of 6.37$\times$. Table~\ref{tab:speed-up} summarizes the overall speedup achieved by our approach compared with the baselines for ranks between 16-256.




\begin{figure*}
    \centering
    \subfigure[Nell-1]{\includegraphics[width=0.18\textwidth]{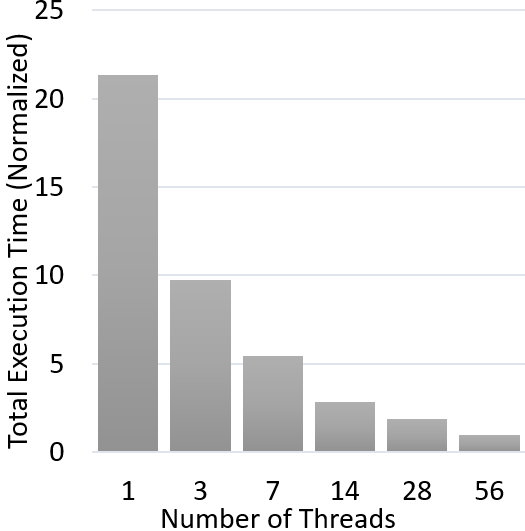}} 
    \subfigure[Nell-2]{\includegraphics[width=0.21\textwidth]{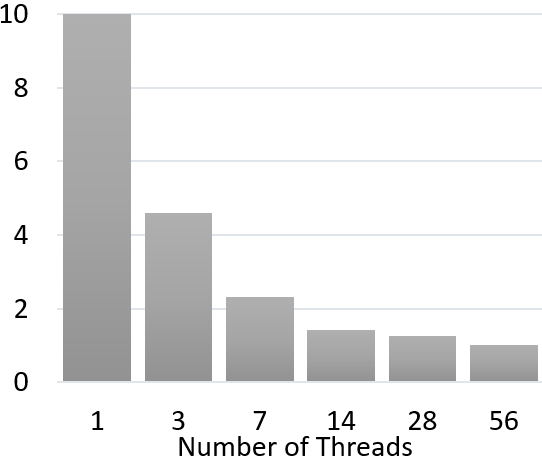}}
    \subfigure[Flickr]{\includegraphics[width=0.19\textwidth]{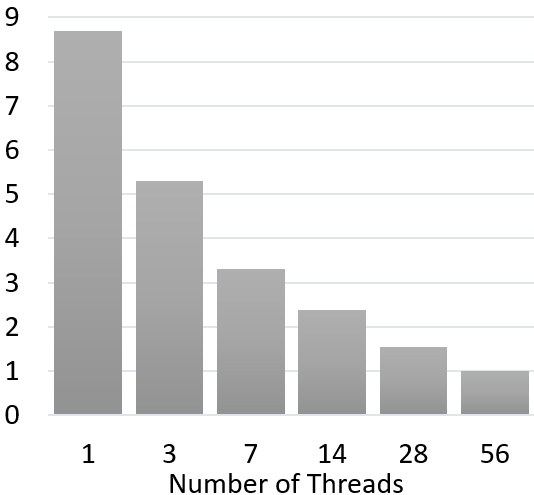}}
    \subfigure[Delicious]{\includegraphics[width=0.18\textwidth]{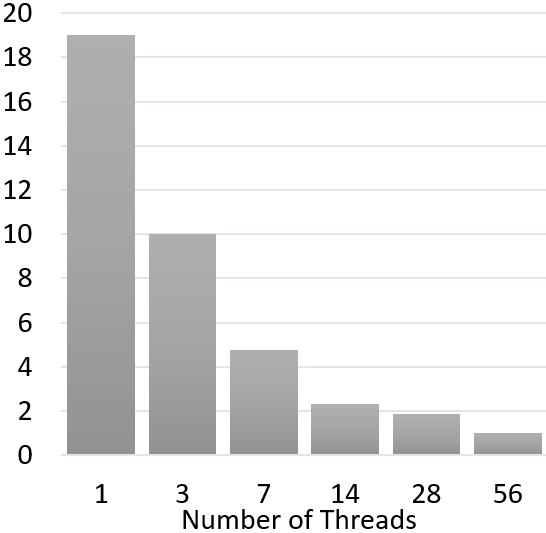}}
    \subfigure[Vast]{\includegraphics[width=0.19\textwidth]{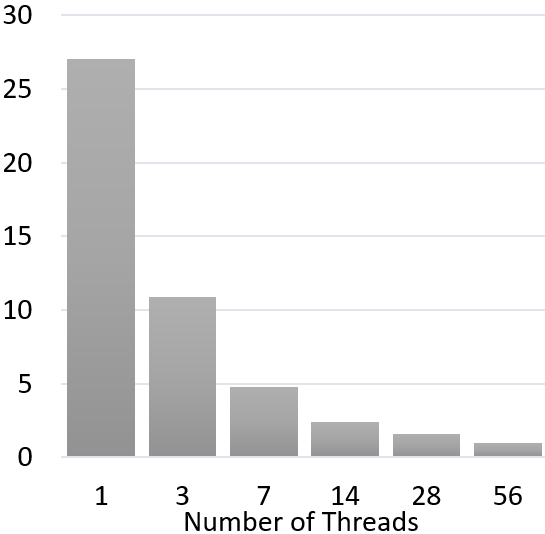}}
    \vspace{-4mm}
    \caption{Scalability with w.r.t. the number of threads on Intel platform ($R = 16$)}
    \label{thread_scalability}
    \vspace{-5mm}
\end{figure*}

\subsection{Impact of Super-shard Scheduling}\label{sec_ss_schedule_res}
\begin{wrapfigure}{r}{0.25\textwidth}
  \begin{center}
    \includegraphics[width=0.25\textwidth]{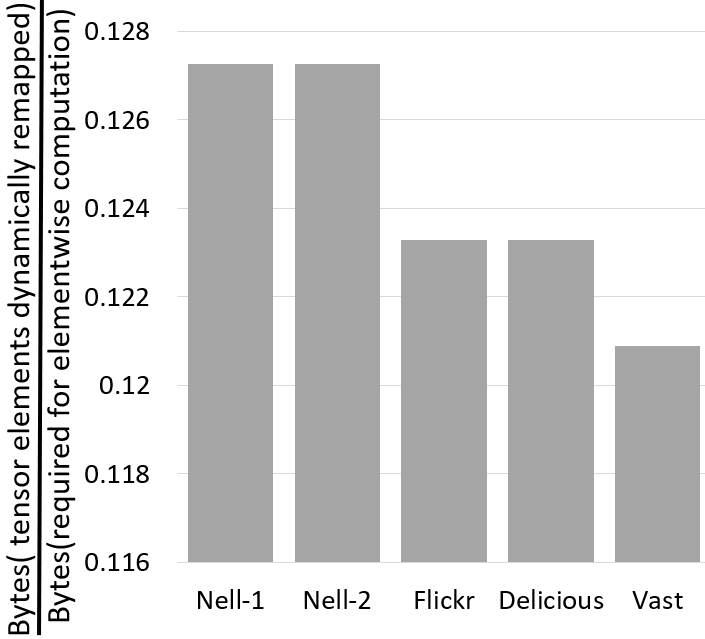}
  \end{center}
  \vspace{-3mm}
  \caption{Total data remapped dynamically vs. total data communicated during the elementwise computations of spMTTKRP}
\vspace{-5mm}
\label{impact_remapped_bits}
\end{wrapfigure}
Section~\ref{load_balancing} presents the scheduling strategy we used in this work to distribute the super-shards among CPU threads. Here, the objective is to distribute the nonzero tensor elements equally among the threads for execution. Our scheduling strategy guarantees the maximum number of nonzero tensor elements scheduled for a thread does not exceed 4/3 times the optimal load in each mode (see Section~\ref{shard_scheduling}).

A state-of-the-art approach to scheduling super-shards among threads in each mode is to distribute the super-shards in block-cyclic manner~\cite{prylli1996efficient}.

Figure~\ref{impact_ss_schedule} compares the state-of-the-art and our proposed method in Section~\ref{shard_scheduling}. The experiments are conducted on the Intel Xeon platform. We also use the same FLYCOO tensor partitioning parameters and memory layout for both cases. Our proposed scheduling strategy achieves 1.1$\times$ to 14.2$\times$ speedups on sparse tensors w.r.t. total execution time.

\vspace{-2mm}
\subsection{Cost of Tensor Remapping}\label{sec_tensor_remap_res}

As shown in Figure~\ref{impact_remapped_bits}, the overall volume of dynamically remapped data during the computation (in all tensors) is consistently below $15\%$ of the total data communicated during the elementwise computations of spMTTKRP. Hence the dynamic tensor remapping does not significantly contribute to the total memory traffic. We determine the volume of data transmitted in various data types, such as tensor elements and factor matrices, by incorporating data element counters into the source code. These counters allow us to measure and track the amount of data communicated during the execution of the program.

\subsection {Scalability}


Figure~\ref{thread_scalability} shows the execution time of the tensors as the number of threads is varied for $R = 16$. Note that execution time is normalized, and the x-axis is in log scale. Utilizing the 56 CPU threads leads to an 8.5$\times$ - 21$\times$ reduction in the overall execution time compared with the single thread implementation. As shown in the roofline model (see Section~\ref{sec_roofline}), the spMTTKRP operation is a memory-bound computation. Hence the speedup saturates as the number of CPU threads is increased. Similar scalability is observed for $R = 32, 64, 128$, and $256$.

\begin{figure}[ht]
\vspace{-5mm}
\centering
\includegraphics[width=0.7\linewidth]{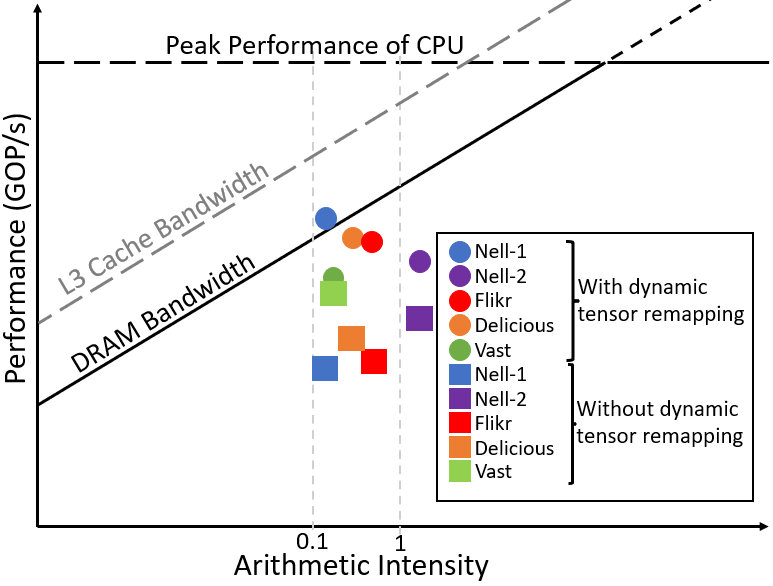}
  \vspace{-2mm}
  \caption{Roofline analysis on Intel platform}
  \label{roofline_m}
  \vspace{-6mm}
\end{figure}

\subsection{Roofline Analysis}\label{sec_roofline}

Figure~\ref{roofline_m} illustrates the roofline model~\cite{10.5555/1999263}, generated using Intel Advisor~\cite{o2017intel} for the datasets on the Intel CPU. The arithmetic intensity of the elementwise spMTTKRP computation (see Section~\ref{sec_elementwise_compute}) is used as the x-axis.

We consider two distinct cases to evaluate the impact of dynamic tensor remapping: Case 1 involves executing Dynasor with dynamic tensor remapping across all modes, while Case 2 involves performing elementwise computation without employing dynamic tensor remapping. In Case 2, we organize the tensor based on shard ids in each mode and perform elementwise computation across all modes without dynamic remapping. For example, for a 3-mode input tensor, we organize the tensor-based shards of mode 0 and perform elementwise computation for modes 0, 1, and 2. This process is repeated after ordering the tensor for all the modes, resulting in multiple roofline values. The best performance value (y-axis) among those is reported in Figure~\ref{roofline_m} as ``without dynamic tensor remapping".

As depicted in Figure~\ref{roofline_m}, including dynamic tensor remapping significantly enhances the performance for all the tensors, compared to the scenario where only the elementwise computation is performed without dynamic tensor remapping.

\subsection{Impact of Factor Matrix Rank}




\begin{wrapfigure}{r}{0.25\textwidth}
  \begin{center}
    \includegraphics[width=0.25\textwidth]{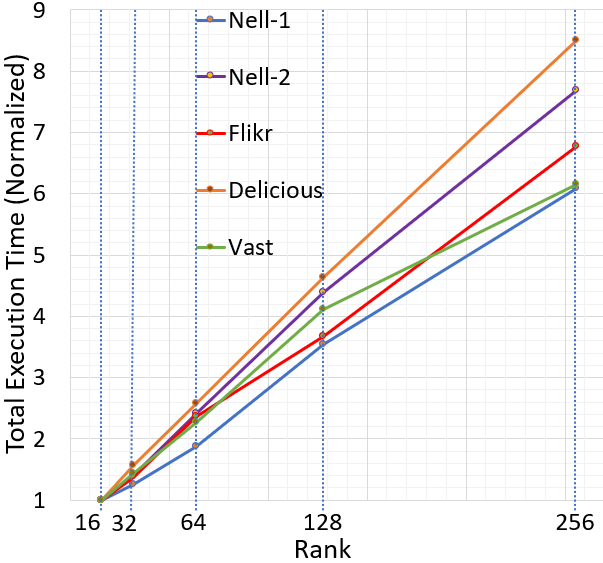}
  \end{center}
\vspace{-5mm}
  \caption{Impact of rank (\textit{R})}
\label{impact_rank}
\vspace{-3mm}
\end{wrapfigure}

The total execution time as the rank of the factor matrices ($R$) is varied, depicted in Figure~\ref{impact_rank}. Loading the input factor matrices dominates the external memory traffic generated in Algorithm~\ref{parallel_alg}. Using the same notation as Section~\ref{Data_partitioning}, the total number of the input factor matrix accesses is proportional to $N \times (N-1) \times |T| \times R$. Consequently, the total CPU memory traffic is proportional to $R$. Since spMTTKRP is a memory-bound computation for sparse tensors (as highlighted in Section~\ref{sec_roofline}), the total execution time is proportional to $R$.





\begin{figure}
    \centering
    \subfigure[Nell-1]{\includegraphics[width=0.40\textwidth]{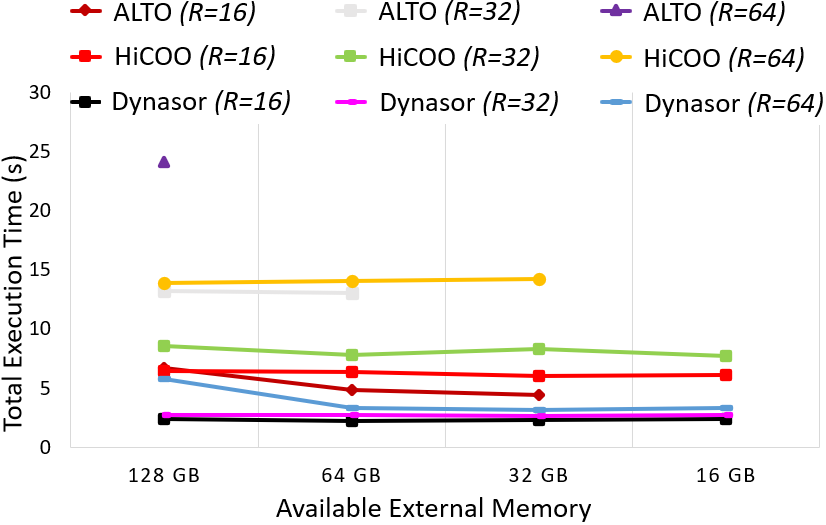}} 
    \vspace{3mm}
    \subfigure[Flickr]{\includegraphics[width=0.40\textwidth]{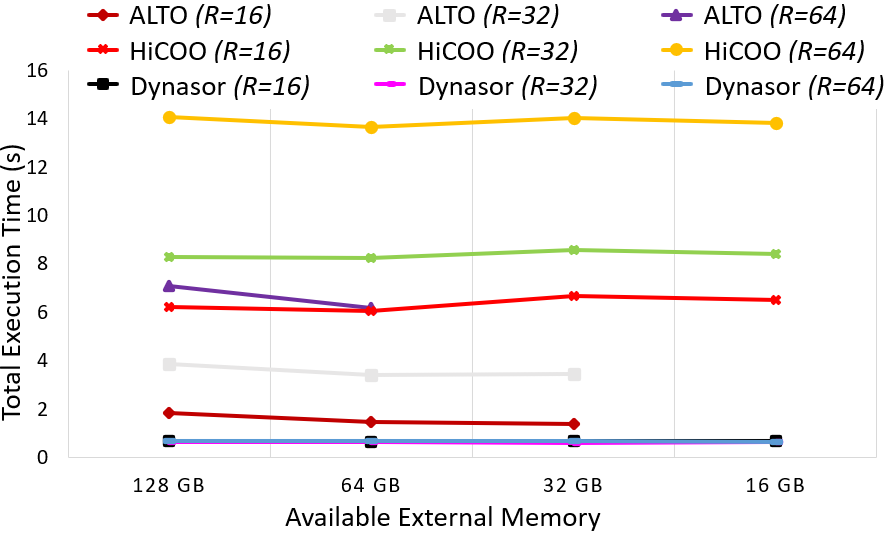}} 
    \vspace{-5mm}
    \caption{Impact of external memory size. Missing data points indicates out-of-memory due to memory explosion}
    \label{fig:lim_mem}
    \vspace{-5mm}
\end{figure}
\vspace{-2mm}
\subsection{Impact of External Memory Size}

Despite the possibility of compressing the total size of the tensor using various tensor formats, excessive generation of intermediate values during runtime can lead to memory explosion. This can result in an out-of-memory error during execution. To illustrate the resilience of Dynasor to the memory explosion, we use 2 of the largest tensors, Nell-1 and Flickr (similar results have been observed in other datasets).

Figure~\ref{fig:lim_mem} (a) and Figure~\ref{fig:lim_mem} (b) show the total execution time of Nell-1 and Flickr when executed on limited CPU external memory. In the experiments, we varied the rank of the factor matrices from 16 to 64. 
For Nell-1 and Flickr, the total external memory requirement to store the tensor memory layout (i.e., including additional space to store remapped tensor), factor matrices, and other metadata stays between 10 GB - 15.6 GB for $R$ values of 16, 32, and 64. Consequently, we vary the size of the CPU external memory (of the Intel platform) between 128 GB (the total available memory) and 16 GB using the OS settings for memory management.



Despite using the low-rank factor matrices ($R = 16$), ALTO cannot perform spMTTKRP with 16 GB of external memory. ALTO also incurs memory explosion as the rank increases, making it impractical for higher ranks. The memory explosion occurs due to excessive intermediate values generated during the execution time. In contrast, HiCOO can be executed in limited memory for low ranks (e.g., $R = 16$). However, ALTO and HiCOO fail to complete the execution using limited external memory as the rank increases (e.g., $R = 64$).

We can perform spMTTKRP in all the cases while achieving minimum execution time. Furthermore, the execution time remains almost the same as the external memory size is decreased for both datasets.

\subsection{Preprocessing Time}

\begin{wrapfigure}{r}{0.28\textwidth}
  \begin{center}
    \includegraphics[width=0.28\textwidth]{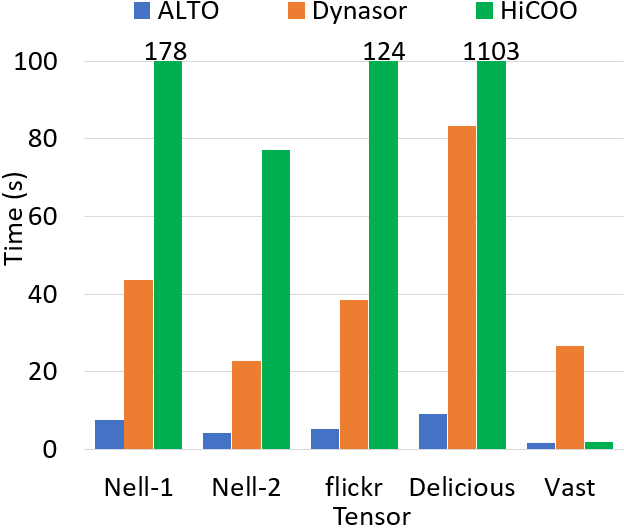}
  \end{center}
  \caption{Preprocessing time on Intel platform}
  \label{tensor_formation}
  \vspace{-5mm}
\end{wrapfigure}

The preprocessing of an input tensor comprises three stages: (1) super-shard generation for each mode, (2) Z-Morton ordering of the super-shards, and (3) shard generation using the super-shards. We have parallelized the preprocessing using OpenMP and Boost library~\cite{schaling2014boost}. Although the focus of our paper is not on the preprocessing time, Figure~\ref{tensor_formation} compares the preprocessing time across various baselines. We use the same Intel Xeon platform (see Section~\ref{sec_platform}) for all the implementations.
The preprocessing time of Dynasor is considerably shorter than that of HiCOO for datasets that contain a large number of indices along each mode. This distinction arises because the partitioning scheme employed by Dynasor focuses solely on the nonzero elements, whereas the HiCOO partitioning scheme operates across the entire index space, encompassing all potential combinations of index values across all the modes.
For tensors with a small number of mode indices but a large number of nonzeros (e.g., Vast), the HiCOO format outperforms our approach in preprocessing time. Furthermore, the ALTO format generation performs faster than Dynasor due to implementation inefficiencies in tensor element ordering based on Z-Morton. We intend to address this issue in future work by further optimizing our preprocessing implementation.
\section{Conclusion and Future Work}
This paper presented a novel parallel algorithm for spMTTKRP across all the modes of an input tensor on multi-core CPUs. We reduced the total execution time of spMTTKRP by employing a parallel algorithm that enables concurrent processing of independent partitions of the input tensor. Furthermore, our algorithm reduces the intermediate values being communicated to external memory. The experimental results demonstrate that our work achieves a geometric mean of 6.37$\times$ speedup in execution time compared with the state-of-the-art CPU implementations across widely-used real-world sparse tensor datasets.

Our future work focuses on adapting the parallel algorithm and FLYCOO tensor format for emerging heterogeneous systems and massively parallel computing platforms. We plan to utilize massively parallel computing platforms, including GPU, to perform spMTTKRP.
\section*{Acknowledgment}
This work is supported by the National Science Foundation (NSF) under grants OAC-2209563, CNS-2009057 and in part by DEVCOM Army Research Lab under grant W911NF2220159. \\
We also acknowledge Jiajia Li (North Carolina State University), Ahmed E. Helal (Intel Labs), and Fabrizio Petrini (Intel Labs) for their support in setting up the baseline experiments.
\\ Distribution Statement A: Approved for public release. Distribution is unlimited.

\bibliographystyle{IEEEtran}
\bibliography{reference}

\end{document}